\documentclass[oldversion]{aa} % for a referee version

\usepackage{graphicx}

\usepackage[varg]{txfonts}

\begin{document}

   \title{Molecular gas in NUclei of GAlaxies (NUGA)}

   \subtitle{XI. A complete gravity torque map of NGC~4579: new clues on bar evolution\thanks{Based on observations carried out with the IRAM Plateau de Bure Interferometer. 
IRAM is supported by INSU/CNRS (France), MPG (Germany) and IGN (Spain)}}

   \author{S.~Garc\'{\i}a-Burillo\inst{1}
		\and
	   S.~Fern\'{a}ndez-Garc\'{\i}a\inst{1}
		\and
	   F.~Combes\inst{2}
		\and
	   L.~K. Hunt\inst{3}
		\and
	   S.~Haan\inst{4}
		\and
	   E.~Schinnerer\inst{4}
		  \and
	   F.~Boone\inst{2}
		  \and
	   M.~Krips\inst{5}
		  \and
	   I. M\'arquez\inst{6}}

   \offprints{S.~Garc\'{\i}a-Burillo}

   \institute{Observatorio Astron\'omico Nacional (OAN)-Observatorio de Madrid,
Alfonso XII, 3, 28014-Madrid, Spain \\
			  \email{s.gburillo@oan.es, s.fernandez@oan.es}
	\and
		Observatoire de Paris, LERMA, 61 Av. de l'Observatoire, 75014-Paris, France \\
			 \email{francoise.combes@obspm.fr, frederic.boone@obspm.fr}
	\and
		 Istituto di Radioastronomia/CNR, Sez. Firenze, Largo Enrico Fermi, 5, 50125-Firenze, Italy \\
			 \email{hunt@arcetri.astro.it}
	\and
	 %National Radio Astronomy Observatory, P.O. Box 0, Socorro, NM87801, USA\\
	 Max-Planck-Institut f\"ur Astronomie, K\"onigstuhl, 17, 69117-Heidelberg, Germany \\
			 \email{haan@mpia.de, schinner@mpia-hd.mpg.de} %schinner@mpia.de
	\and
		Center for Astrophysics, SMA project, 60 Garden
Street, MS 78 Cambridge, MA-02138, USA \\
			 \email{mkrips@cfa.harvard.edu}
	 \and
		Instituto de Astrof\'{\i}isica de Andaluc\'{\i}a (CSIC), Apdo 3004, 18080-Granada, Spain \\ 
			\email{isabel@iaa.es}
}
	\date{Received ---; accepted ----}

\abstract{

In this paper we create a complete gravity torque map of the disk of the LINER/Seyfert 1.9 galaxy NGC\,4579. We quantify the efficiency of angular momentum transport and search for signatures of secular evolution in the fueling process from $r$\,$\sim$\,15~kpc down to the inner $r$\,$\sim$50\,pc around the Active Galactic Nucleus (AGN). We use both the 1--0 and 2--1 line maps of $^{12}$CO obtained with the Plateau de Bure Interferometer (PdBI) as part of the {\it NUclei of Galaxies}--(NUGA)--project. 
The CO(1--0) and CO(2--1) PdBI maps, at 2.0$\arcsec\times$1.3$\arcsec$ and 1.0$\arcsec\times$0.6$\arcsec$ resolution respectively, include short spacing correction provided by IRAM-30m data. We derive the stellar potential from a NIR (K band) wide field image of the galaxy. The $K$-band image, which reveals a stellar bar, together with a high resolution HI map of NGC\,4579 obtained with the Very Large Array (VLA), allow us to extend the gravity torque analysis to the outer disk. Most of the molecular gas mass in the inner $r\leq$2~kpc disk is distributed in two  spiral arms, an outer ring, and a central lopsided disk. The bulk of the gas response traced by the CO PdBI maps follows the expected gas flow pattern induced by the bar potential  in the presence of two Inner Lindblad Resonances (ILR) at $r\sim$500~pc and $r\sim$1.3~kpc . %Assuming that the corotation resonance lies at the end of the bar ($r_{CR}\sim$6~kpc), we estimate a bar pattern speed $\Omega_{BAR}\sim$50~km~s$^{-1}$kpc$^{-1}$ and two ILRs at $r\sim$500~pc (inner ILR=iILR) and $r\sim$1.3~kpc (outer ILR=oILR).
We also detect an oval distortion in the inner $r\sim$200~pc of the $K$-band image. The oval is not aligned with the large-scale bar, a signature of dynamical decoupling. The morphology of the outer disk suggests that the neutral gas is currently piling up in a pseudo-ring formed by two winding spiral arms that are morphologically decoupled from the bar structure. The pseudo-ring is located inside the bar corotation ($r_{CR}\sim$6~kpc) and close to the predicted position of the Ultra Harmonic Resonance (UHR) ($r_{UHR}\sim$3.8~kpc). 

The derived gravity torque budget in NGC~4579 shows that the fueling process is at work on different spatial scales in the disk. In the outer disk ($r\geq$2~kpc), the decoupling of the spiral allows the gas to efficiently populate the UHR region, and thus produce net gas inflow on intermediate scales. The corotation barrier seems to be overcome due to secular evolution processes. The gas in the inner disk ($r\leq$2~kpc) is efficiently funneled by gravity torques down to $r\sim$300~pc.  Closer to the AGN ($r<$200~pc), gas feels negative torques due to the combined action of the large-scale bar and the inner oval. The two $m=2$ modes act in concert to produce net gas inflow down to $r\sim$50~pc, providing a clear {\it smoking gun} evidence of fueling with associated short dynamical time-scales ($\sim$1--3 rotation periods).

}

%
%________________________________________________________________
   \maketitle

\section{Introduction}\label{introduction}

Nuclear activity in galaxies is understood as the result of the feeding of supermassive black holes (SMBH).
SMBH are suspected to be a common component in most, perhaps all, galaxies with a significantly massive bulge (e.g., see review by Ferrarese \& Ford~\cite{fer05} and references therein). Active Galactic Nuclei (AGN) must be fed with material coming from the disk of the host galaxy. The supplying gas, lying originally far away from the gravitational influence of the black hole, must lose virtually all of its angular momentum in the fueling process. While for high luminosity AGNs, a good correlation between the presence of $\sim$kpc-scale non-axisymmetric perturbations and the existence of activity has been found (e.g., Hutchings \& Neff~\cite{hut92}), the case for a similar correlation in low luminosity AGNs (LLAGNs) is weak if any (e.g., Mulchaey \& Regan \cite{mul97}; Knapen et al.~\cite{kna00}; M\'arquez et al.~\cite{mar00}; Combes~\cite{com03}). The search for a universal mechanism for AGN feeding in LLAGNs is probably complicated by the fact that the AGN duty cycle ($\sim$10$^{6-7}$ years) might be shorter than the lifetime of the feeding mechanism itself (e.g., Wada~\cite{wad04}; Hopkins \& Hernquist~\cite{hop06}; King \& Pringle~\cite{kin07}). In spite of all the theoretical and observational efforts, finding a solution for the LLAGN fueling problem has thus far remained an elusive aim (e.g., see review by Martini~\cite{mar04}).

The study of the content, distribution and kinematics of interstellar gas is a key to understand the origin and maintenance of nuclear activity in galaxies. The processes involved in AGN fueling encompass a wide range of scales, both spatial and temporal, which have to be studied. Probing the gas flow from the outer disk down to the central engine of an AGN host, requires the use of specific tracers of the interstellar medium adapted to follow the change of phase of the gas as a function of radius. As most of the neutral gas in galactic nuclei is in the molecular phase, low-J rotational lines of carbon monoxide (CO) are the best choice to undertake high-spatial resolution ($\simeq$1$\arcsec$) interferometer mapping of the central kiloparsec disks of AGNs. On the other hand, the 21cm line emission of atomic hydrogen (HI) is best suited to trace the distribution and kinematics of neutral gas in the outer disk reservoirs of AGNs ($r$\,$\geq$5~kpc) with moderate spatial resolution ($\simeq$10--20$\arcsec$) using an interferometer. Besides providing a handle on the content and distribution of the gas, CO and HI interferometer maps give a sharp view of the gas kinematics. The combined information extracted from CO and HI is essential to characterize gravitational instabilities and constrain theoretical models of the gas flows at all spatial scales in galaxy disks.

The NUclei of GAlaxies--NUGA--project, described by Garc\'{\i}a-Burillo et
al.~(\cite{gb03a,gb03b}), is the first high-resolution ($\sim$0.5$^{\prime\prime}$--1$^{\prime\prime}$) $^{12}$CO survey of 12 nearby ($D$=4--40~Mpc) LLAGNs including the full sequence of activity types (Seyferts, LINERs and transition objects). Observations, carried out with the IRAM Plateau de Bure Interferometer (PdBI), have been completed early 2004. NUGA surpasses in both spatial resolution (10--100~pc) and sensitivity (3$\sigma$--detection limit$\simeq$a few~10$^{5-6}M_{\sun}$) the previous surveys of nearby AGN conducted at the Owens Valley Radio Observatory--OVRO (Baker~\cite{bak00}; Jogee et al.~\cite{jog01}) and at the Nobeyama Radio Observatory--NRO (Kohno et al.~\cite{koh01}). NUGA maps allow us to probe the gas flows at critical spatial scales ($<$100~pc) where secondary modes embedded in the kpc-scale perturbations are expected to take over in the fueling process. A bottom line result of NUGA is the identification of a wide range of morphologies in the central kpc-disks of these LLAGNs. This includes one-arm spirals or $m=1$ instabilities (NGC~4826: Garc\'{\i}a-Burillo et al.~\cite{gb03b}; NGC~3718: Krips et al.~\cite{kri05}), symmetric rings (NGC~7217: Combes et al.~\cite{com04}; NGC~3147: Casasola et al.~\cite{cas07}), as well as gas bars and two-arm spirals (NGC~4569: Boone et al.~\cite{boo07}; NGC~2782: Hunt et al.~\cite{hun08}; NGC~6574: Lindt-Krieg et 
al.~\cite{lin08}). As such, this result is suggestive of an evolutionary scenario where several mechanisms cooperate to feed the central engines of LLAGNs. We have performed a detailed case-by-case study of the distribution and kinematics of molecular gas in the galaxies of our sample, and interpreted these in terms of evidence of ongoing feeding. The gas response to the stellar potential is characterized with the help of high-resolution optical and NIR images of the galaxies. We adopt this in-depth approach to take full advantage of the high quality of NUGA maps.

In a previous pilot study, Garc\'{\i}a-Burillo et al.~(\cite{gb05}) (hereafter GB05) analyzed the efficiency of stellar gravity torques to drain the gas angular momentum in the central kiloparsec of four NUGA targets: NGC\,4321, NGC\,4826, NGC\,4579 and NGC\,6951. Results of this analysis indicate that, paradoxically, feeding due to the stellar potential is presently quenched close to the AGNs. Due to the inhibiting action of bars (NGC\,4321, NGC\,4579) or weak oval perturbations (NGC\,4826, NGC\,6951), the derived gravity torque budget is seen to be positive inside $r$\,$\sim$150-200~pc in the four LLAGNs analyzed in this paper. This situation is consistent with the theoretical picture where gas gets trapped in the Inner Lindblad Resonance (ILR) of large-scale bars. ILRs favor the feeding of a starburst in nuclear rings but can halt the gas flow inwards (Combes~\cite{com88}; Regan \& Teuben~\cite{reg04}). GB05 speculate that the agent responsible for driving inflow at smaller radii may be transient (lifetime$\leq$10$^6$~yrs); finding a smoking gun evidence of fueling would be thus difficult (see however the case of NGC\,2782 analyzed by Hunt et al.~\cite{hun08}). As an alternative explanation,  GB05 estimate on a case-by-case basis that the gravity torque barrier associated with the ILRs in these galaxies could be overcome by viscosity. In this scenario, gravity torques and viscosity coordinate efforts to produce recurrent episodes of activity during the typical lifetime of any galaxy. The weakening of bars, triggered by the radial re-distribution of gas and the ensuing angular momentum exchanges (Bournaud \& Combes~\cite{bou02}), may allow AGN feeding episodes to occur. This would account for the presence of filamentary dusty structure connecting nuclear rings to the central engines of some LLAGNs (e.g., Peeples \& Martini~\cite{pee06}).

%%%%%%%%%%%%%%%%%%%%%%%%%%%%%%%%%%%%%%%%%%%%%%%%%%%%%%%%%%%%%%%%%%%%%%%%%%%%%%%%%%%%%%%%%%%%%%%%%%%%%%%%%%%%%%%%

 \begin{table}
   \caption[]{Observational parameters of NGC~4579.}
   \begin{center}
   \begin{tabular}{lll}
   \hline
   \hline
     Parameter  & Value & Reference \\
   \hline
   $\alpha_{J2000}$ (dynamical center) & 12$^{h}$37$^{m}$43.52$^{s}$ & This work \\
   $\delta_{J2000}$ (dynamical center) & 11$^{\circ}$49$^{\prime}$05.5$\arcsec$ & This work \\
   V$_{hel}$ & 1466 km s$^{-1}$ & This work \\
   RC3 Type & SAB(rs)b & NED \\
   AGN Type & LINER/Sy1.9 & NED \\
   t Type & 2.8 & LEDA \\
   Inclination & 36$^{\circ}$ & GB05 \\
   Position Angle & 95$^{\circ}$ & GB05 \\
   Distance & 19.8 Mpc (1$^{\prime\prime}$=97 pc) & GB05 \\
   M$_{B}$ & -21.68 mag & LEDA \\
   M(HI) & 1.7 $\times 10^{8}$ M$_{\odot}$ & Haan et al.~(\cite{haa08a}) \\
   M(H$_{2}$) & 5 $\times 10^{8}$ M$_{\odot}$ & This work \\
   L$_{FIR}$ & 9.6 $\times 10^{9}$ L$_{\odot}$ & Sanders et al.~(\cite{san03}) \\
   \hline
   \hline
  \end{tabular}
   \label{Table1}
   \end{center}
\end{table}

%
% The recurrence of activity in galaxies would be thus indirectly linked to that of the bar instabilities although the 
% corresponding peaks of activity and bar phases would not necessarily coincide, as observed in NUGA galaxies.
%
%%%%%%%%%%%%%%%%%%%%%%%%%%%%%%%%%%%%%%%%%%%%%%%%%%%%%%%%%%%%%%%%%%%%%%%%%%%%%%%%%%%%%%%%%%%%%%%%%%%%%%%%%%%%%%%%%%%%%%%%

%%%%%%%%%%%%%%%%%%%%%%%%%%%%%%%%%%%%%%%%%%%%%%%%%%%%%%%%%%%%%%%%%%%%%%%%%%%%%%%%%%%%%%%%%%%%%%%%%%%%%%%%%%%%%%%%

\begin{figure*}[tbh!]
 \centering
 \includegraphics[width=8.5cm, angle=-90]{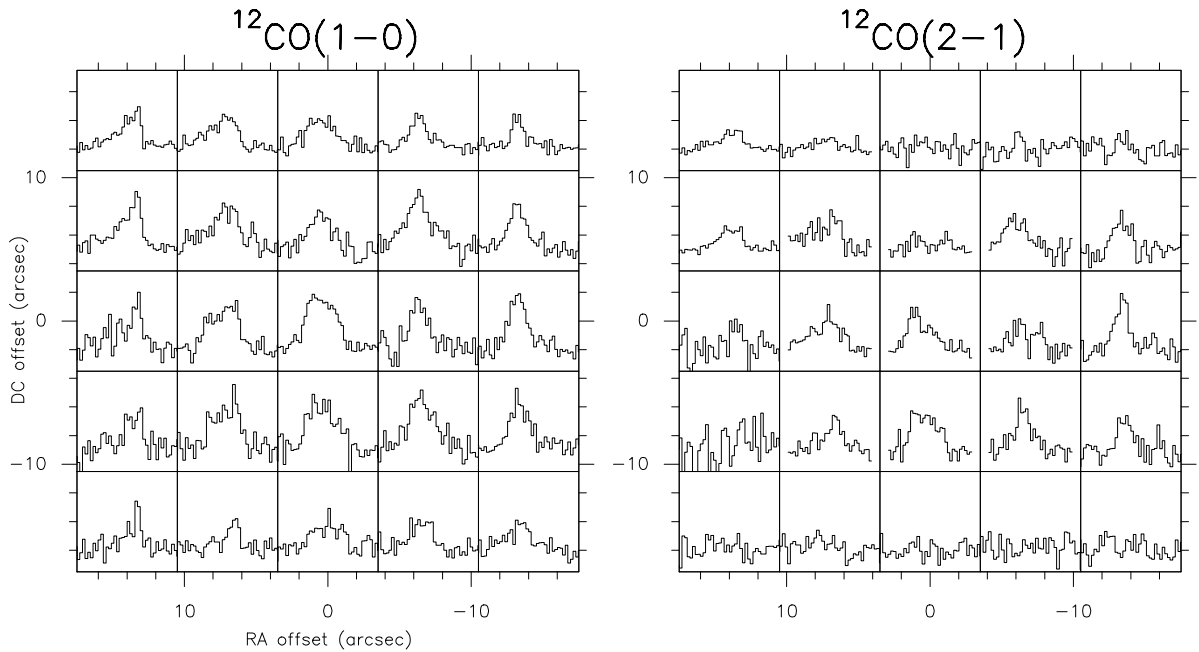}
  \caption{{\bf a)} ({\it Left panel})~CO(1--0) line emission profiles observed with the IRAM 30m telescope in NGC~4579. We have observed with a 5$\times$5-point grid and a 7$\arcsec$ spacing. Offsets are relative to the AGN. Velocity scale goes from
--400~km~s$^{-1}$ to 400~km~s$^{-1}$ with respect to $v$=$v_{0}$=1520~km~s$^{-1}$ in LSR scale, and temperature scale (in T$_a^*$) goes from --30 to 100~mK. {\bf b)} ({\it Right panel}) Same as {\bf a)} but for the ~CO(2--1) line. Same velocity scale as in {\bf a)}; temperature scale from --40 to 150~mK. All spectra have been smoothed to a resolution of 10~km~s$^{-1}$. }
\label{fig:30m-spectra} 
\end{figure*}

%%%%%%%%%%%%%%%%%%%%%%%%%%%%%%%%%%%%%%%%%%%%%%%%%%%%%%%%%%%%%%%%%%%%%%%%%%%%%%%%%%%%%%%%%%%%%%%%%%%%%%%%%%%%%%%%%%%%%%%%

NGC\,4579 is a SAB(rs)b galaxy classified as an intermediate type 1 object (LINER/Seyfert 1.9) by Ho et al.~(\cite{ho97}). The AGN nature of NGC\,4579 is confirmed by the detection of broad H$_{\alpha}$ and UV lines (Stauffer~\cite{sta82}; Filippenko \& Sargent~\cite{fil85}; Barth et al.~\cite{bar96, bar01}; Maoz et al.~\cite{mao98}).
It also has an unresolved nuclear hard X-ray (variable) source with a prominent broad Fe K$_{\alpha}$ line (Terashima et al.~\cite{ter00}; Ho et al.~\cite{ho01a}; Eracleous et al.~\cite{era02}; Dewangan et al.~\cite{dew04}). A non-thermal radio continuum source is detected at the position of the AGN (Hummel et al.~\cite{hum87}; Ho \& Ulvestad~\cite{ho01b}; Ulvestad \& Ho~\cite{ulv01}; Krips et al.~\cite{kri07}). Hubble Space Telescope (HST) optical and UV images
resolve the central kiloparsec disk of NGC\,4579 and show a nuclear component surrounded by a highly structured disk (Pogge et al.~\cite{pog00}; Contini~\cite{con04}). The galaxy is part of the Virgo cluster, although neither the morphology nor the kinematics of the HI gas disk, mapped by Haan et al.~(\cite{haa08a}), show any signs of significant stripping at work. 

In this paper we re-determine the gravitational torque budget in the central kiloparsec of 
NGC\,4579 using both the 1--0 and 2--1 line maps of $^{12}$CO obtained inside the NUGA project. The new maps used in this work include the short spacing correction provided by the IRAM-30m data obtained in the two lines of CO, and thus give a more reliable estimate of the implied gas flow time-scales, compared to the first estimate of GB05.  
We derive the stellar potential from a NIR (K band) wide field image of the galaxy disk; this allows us to minimize the residual effects of extinction that were present in the HST $I$-band image that we originally used in the first gravity torque analysis of this galaxy (GB05). Furthermore, the large field-of-view of the $K$-band image, together with the availability of a high resolution and sensitivity HI map of NGC\,4579, recently obtained by Haan et al.~(\cite{haa08a}), allow us to extend the gravity torque analysis to the outer disk of the galaxy. A complete gravity torque map of NGC\,4579, makes it possible to quantify the efficiency of angular momentum transport from $r$\,$\sim$\,a few kpc down to the inner $r$\,$\sim$50\,pc around the AGN and search for signatures of secular evolution in the accretion process. This is a key to test the fueling scenario described by GB05.

We will assume a distance to NGC\,4579 of around 20~Mpc for consistency with Garc\'{\i}a-Burillo et
al.~(\cite{gb05}), which implies 1$\arcsec$\,$\simeq$100\,pc. This is a good compromise between the different values reported in the literature ranging from $D$=16.8~Mpc (Tully \& Fisher~\cite{tul88}) to $D$=22.6~Mpc (LEDA, 2006). Similarly to GB05 we will assume that the inclination angle and the position angle of NGC\,4579's disk are $i$=36$^{\circ}$ and $PA$=95$^{\circ}$, respectively, in rough agreement with previous determinations (e.g., Rubin et al.~\cite{rub99}; Koopmann et al.~\cite{koo01}; Daigle et al.~\cite{dai06}). The basic observational parameters of NGC\,4579 are listed in Table~\ref{Table1}.

We describe in Sect.~\ref{observations} the observations used, including high-resolution CO and HI maps as well as ground-based NIR images of NGC\,4579. Sects.~\ref{AGN} to ~\ref{CO-kinematics} describe 
the distribution and kinematics of molecular gas. We describe the stellar structure and star formation of NGC\,4579 in Sect.~\ref{tracers}. Sect.~\ref{grav} computes from NIR images the gravitational potentials and forces,
and deduces from the combined CO and HI maps the effective torques applied to the gas. From these torques, it is
possible to derive time-scales for gas flows and discuss whether gravity torques alone are
efficient enough to feed the current activity in NGC\,4579. The general implications of these results for the current understanding of AGN feeding are summarized in Sect.~\ref{summary}.
%%%%%%%%%%%%%%%%%%%%%%%%%%%%%%%%%%%%%%%%%%%%%%%%%%%%%%%%%%%%%%%%%%%%%%%%%%%%%%%%%%%%%%%%%%%%%%%%%%%%%%%%%%%%%%%%%%%%

\begin{figure*}[bth!]
 \centering
 \includegraphics[width=16cm]{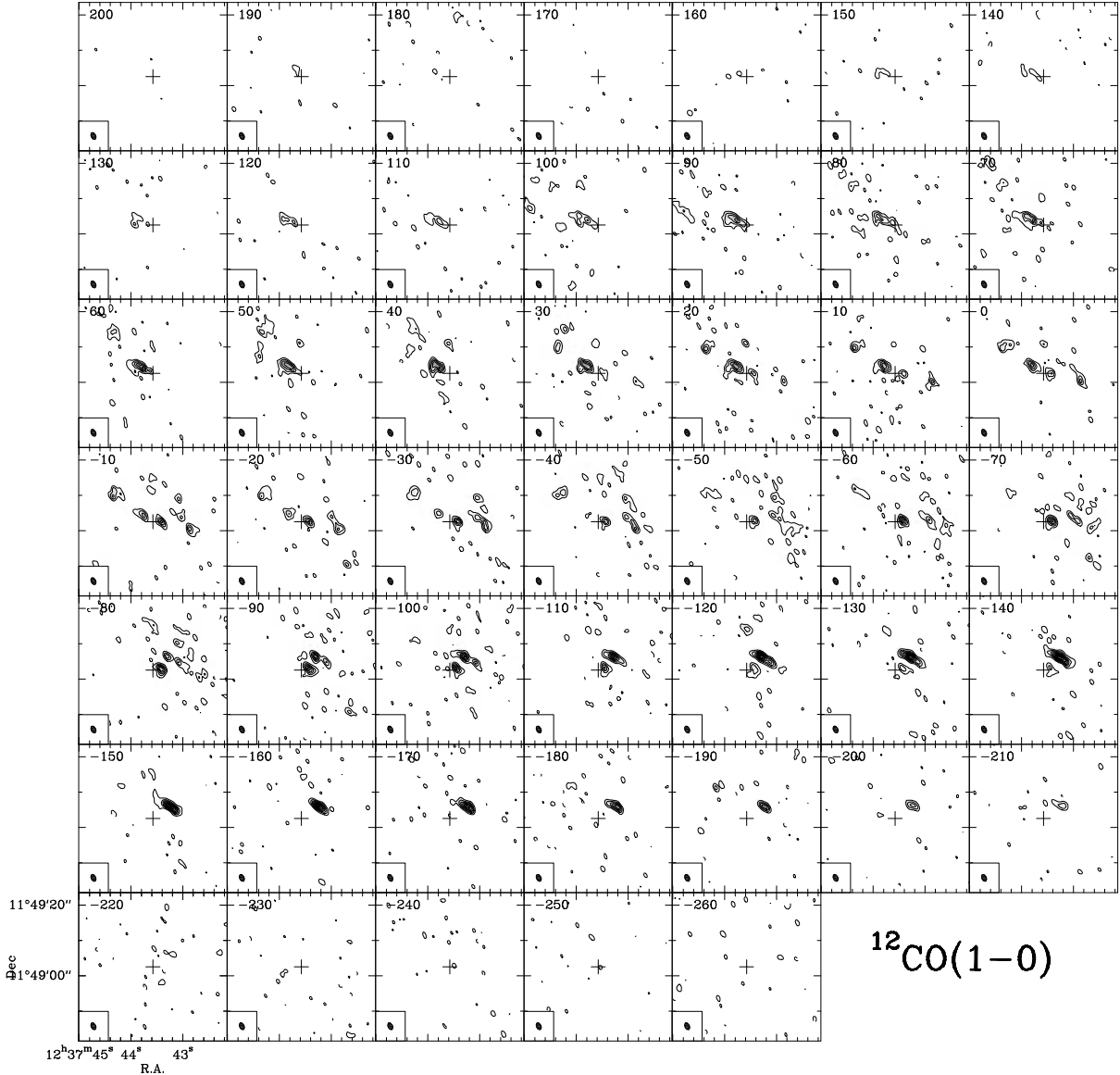}
 \caption{~$^{12}$CO velocity-channel maps observed with the PdBI in the nucleus of NGC~4579 with a spatial resolution of 2.0$\arcsec\times$1.3$\arcsec$ at $PA$=26~$^{\circ}$ (beam is plotted as a filled ellipse in the bottom left corner of each panel). We show a field of view of 42$\arcsec$, i.e.~$\sim$the diameter of the primary beam at 115~GHz. The phase tracking center is indicated by a cross at $\alpha_{2000}$=12$^{h}$37$^{m}$43.58$^{s}$ and $\delta_{2000}$=11$^{\circ}$49$^{\prime}$02.5$\arcsec$. Velocity-channels are displayed from $v$\,=200~km~s$^{-1}$ to $v$\,=--260~km~s$^{-1}$ in steps of 10~km~s$^{-1}$. Velocities are in LSR scale and relative to $v$=$v_{0}$=1520~km~s$^{-1}$. Contour levels are --3$\sigma$, 3$\sigma$ to 24$\sigma$ in steps of 3$\sigma$ where the 1-sigma rms $\sigma$=2.3~mJy~beam$^{-1}$. There are very few pixels with negative fluxes below --3$\sigma$ inside the field-of-view.}
 \label{fig:channels-CO1--0}
\end{figure*}
%%%%%%%%%%%%%%%%%%%%%%%%%%%%%%%%%%%%%%%%%%%%%%%%%%%%%%%%%%%%%%%%%%%%%%%%%%%%%%%%%%%%%%%%%%%%%%%%%%%%%%%%%%%%%%%%%%%%%%%%%
%%%%%%%%%%%%%%%%%%%%%%%%%%%%%%%%%%%%%%%%%%%%%%%%%%%%%%%%%%%%%%%%%%%%%%%%%%%%%%%%%%%%%%%%%%%%%%%%%%%%%%%%%%%%%%%%%%%%%%%%%
\begin{figure*}[bth!]
\centering
     \includegraphics[width=16cm]{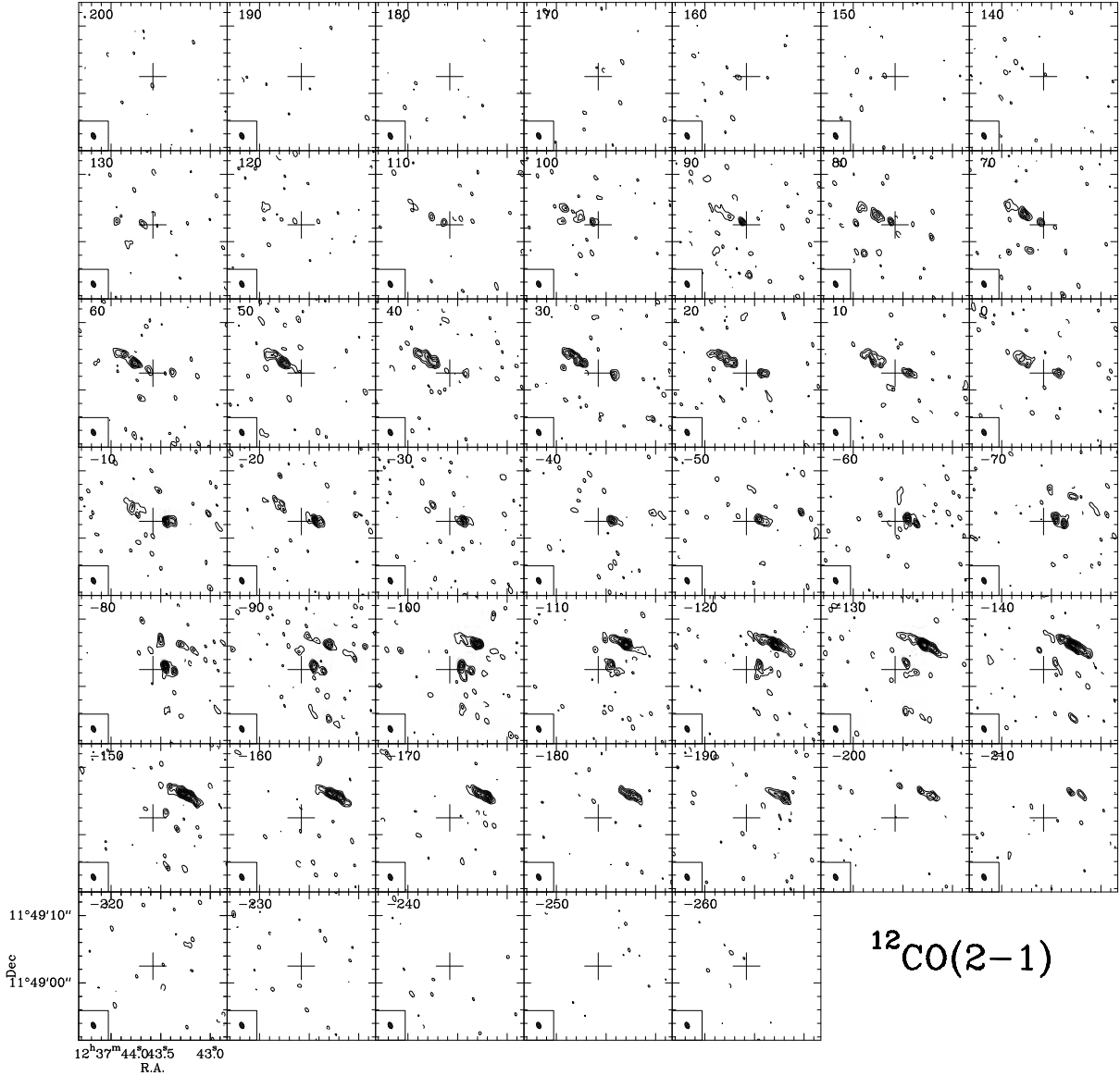}
\caption{~Same as Fig.~1 but for the 2--1 line of CO. Spatial resolution reaches 1.0$\arcsec\times$0.6$\arcsec$ at $PA$=--155$^{\circ}$ (beam is plotted as a filled ellipse in the bottom left corner of each panel). We show a field of view of 21$\arcsec$, i.e. $\sim$diameter of the primary beam at 230 GHz. Velocity-channels are displayed from $v$=200~km~s$^{-1}$ to $v$=--260~km~s$^{-1}$ in steps of 10~km~s$^{-1}$, with same reference as used in Fig. 1. Contour levels are --3$\sigma$, 3$\sigma$ to 19$\sigma$ in steps of 2$\sigma$ where the 1-sigma rms $\sigma$=4.6~mJy~beam$^{-1}$. There are very few pixels with negative fluxes below --3$\sigma$ inside the field-of-view.}
	\label{fig:channels-CO2--1}
\end{figure*}
%%%%%%%%%%%%%%%%%%%%%%%%%%%%%%%%%%%%%%%%%%%%%%%%%%%%%%%%%%%%%%%%%%%%%%%%%%%%%%%%%%%%%%%%%%%%%%%%%%%%%%%%%%%%%%%%%%%%%%%%%

%%%%%%%%%%%%%%%%%%%%%%%%%%%%%%%%%%%%%%%%%%%%%%%%%%%%%%%%%%%%%%%%%%%%%%%%%%%%%%%%%%%%%%%%%%%%%%%%%%%%%%%%%%%%%%%%

\begin{figure*}[tbh!]
 \centering
 \includegraphics[width=18cm]{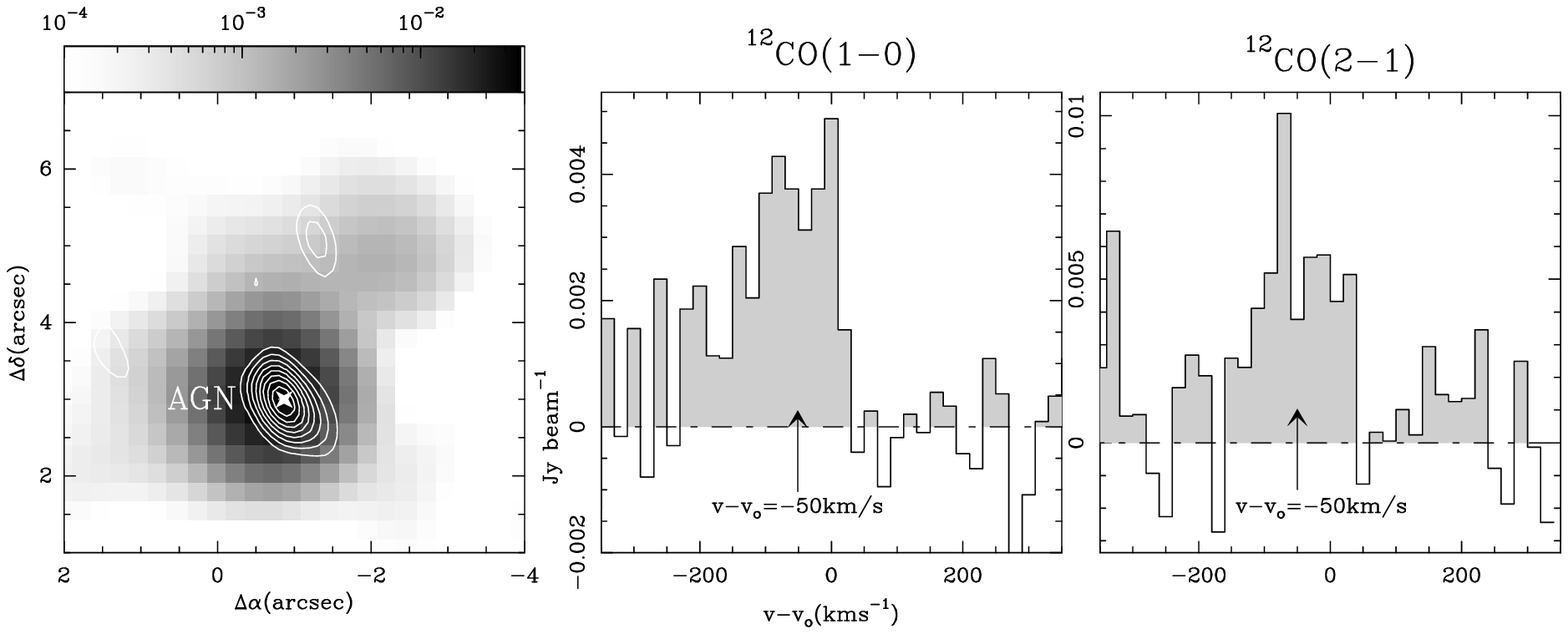}
  \caption{{\bf a)}({\it Left panel})~The continuum map at 1 mm obtained with the PdBI (in contours from 20\% to 100\%, in step of 10\% of the maximum=10~mJy~beam$^{-1}$) is overlaid on the VLA radio-continuum image at 6 cm (in grey scale) taken from Ho \& Ulvestad~(\cite{ho01b}). The AGN is clearly identified with a non-thermal point source. ($\Delta\alpha$,~$\Delta\delta$)--offsets are here relative to the phase tracking center. {\bf b)}({\it Middle panel})~We show the emission associated with the AGN source (torus) observed in the CO(1--0) line. {\bf c)}({\it Right panel})~ The same as {\bf b)} but for CO(2--1), centered around v$_{0}$. The derived systemic velocity $v_{sys}$=$v$--$v_{0}$=--50~km~s~$^{-1}$.}
\label{fig:agn}
\end{figure*}
%%%%%%%%%%%%%%%%%%%%%%%%%%%%%%%%%%%%%%%%%%%%%%%%%%%%%%%%%%%%%%%%%%%%%%%%%%%%%%%%%%%%%%%%%%%%%%%%%%%%%%%%%%%%%%%%

\section{Observations}\label{observations}
\subsection{Interferometer CO maps}

Observations of NGC\,4579 were carried out as
part of the NUGA survey conducted with the PdBI during two years and finished by February 2003. We used the ABCD set of configurations and the six antennas of the array in dual frequency mode to reach the highest spatial resolution ($<$1$^{\prime\prime}$ at the highest
frequency) but also to maximize sensitivity to all spatial frequencies in the maps. 
We observed simultaneously the emission of the J=1--0 and J=2--1 lines of CO in single fields of
sizes 42$^{\prime\prime}$ and 21$^{\prime\prime}$, respectively, centered at $\alpha_{J2000}$=$12^h37^m43.58^s$ 
and $\delta_{J2000}$=$11^{\circ}49'02.5''$. The first results derived from the analysis of the CO(2--1) maps were discussed by GB05. During the observations the spectral correlator was split in two halves centered at the transition rest frequencies corrected for the assumed recession velocity $v_{o}(LSR)$=1520\,km~s$^{-1}$. The correlator configuration covers a bandwidth of 580\,MHz for each line, using four 160\,MHz-wide units with an overlap of 20\,MHz; this is
equivalent to 1510\,km~s$^{-1}$(755\,km~s$^{-1}$) at 115\,GHz (230\,GHz). Visibilities were obtained
using on-source integration times of 20 minutes framed by short ($\sim$\,2\,min) phase and amplitude
calibrations on the nearby quasars 1156+295 and 3C273. The absolute flux scale in our maps was derived to a 10\%
accuracy based on the observations of primary calibrators whose fluxes were determined from a combined set of measurements obtained at the 30m telescope and the PdBI array. The bandpass calibration was carried out using 3C273 and is accurate to better than 5\%. The point source sensitivities derived from emission-free channels of 10~km~s$^{-1}$
width are 2.55\,mJy~beam$^{-1}$ in CO(1--0) and 5.44\,mJy~beam$^{-1}$ in
CO(2--1). Images of the continuum emission of the galaxy at 115\,GHz and 230\,GHz have been obtained by averaging those channels free of line emission. The image reconstruction was done using
standard IRAM/GAG software (Guilloteau \& Lucas~\cite{gui00}). We used natural weighting and no taper to generate the 1--0 line maps with a field of view of 76.8$\arcsec$ and 0.15$''$ sampling; the corresponding synthesized beam is $2.0''\times 1.3''$, $PA$=26$^{\circ}$. We also used natural weighting to generate 2--1 maps with a field of view of
51.2$\arcsec$ and 0.10$''$ sampling; this enables us to achieve a spatial resolution $<$1$''$ 
($1.0''\times 0.6''$, $PA$=25$^{\circ}$). The conversion factors
between Jy\,beam$^{-1}$ and K are 36\,K~Jy$^{-1}$~beam at 115\,GHz, and 40\,K~Jy$^{-1}$~beam at
230\,GHz.

%%%%%%%%%%%%%%%%%%%%%%%%%%%%%%%%%%%%%%%%%%%%%%%%%%%%%%%%%%%%%%%%%%%%%%%%%%%%%%%%%%%%%%%%%%%%%%%%%%%%%%%%%%%%%%%%

\begin{figure}[tbh!]
\centering
    \includegraphics[width=8.5cm]{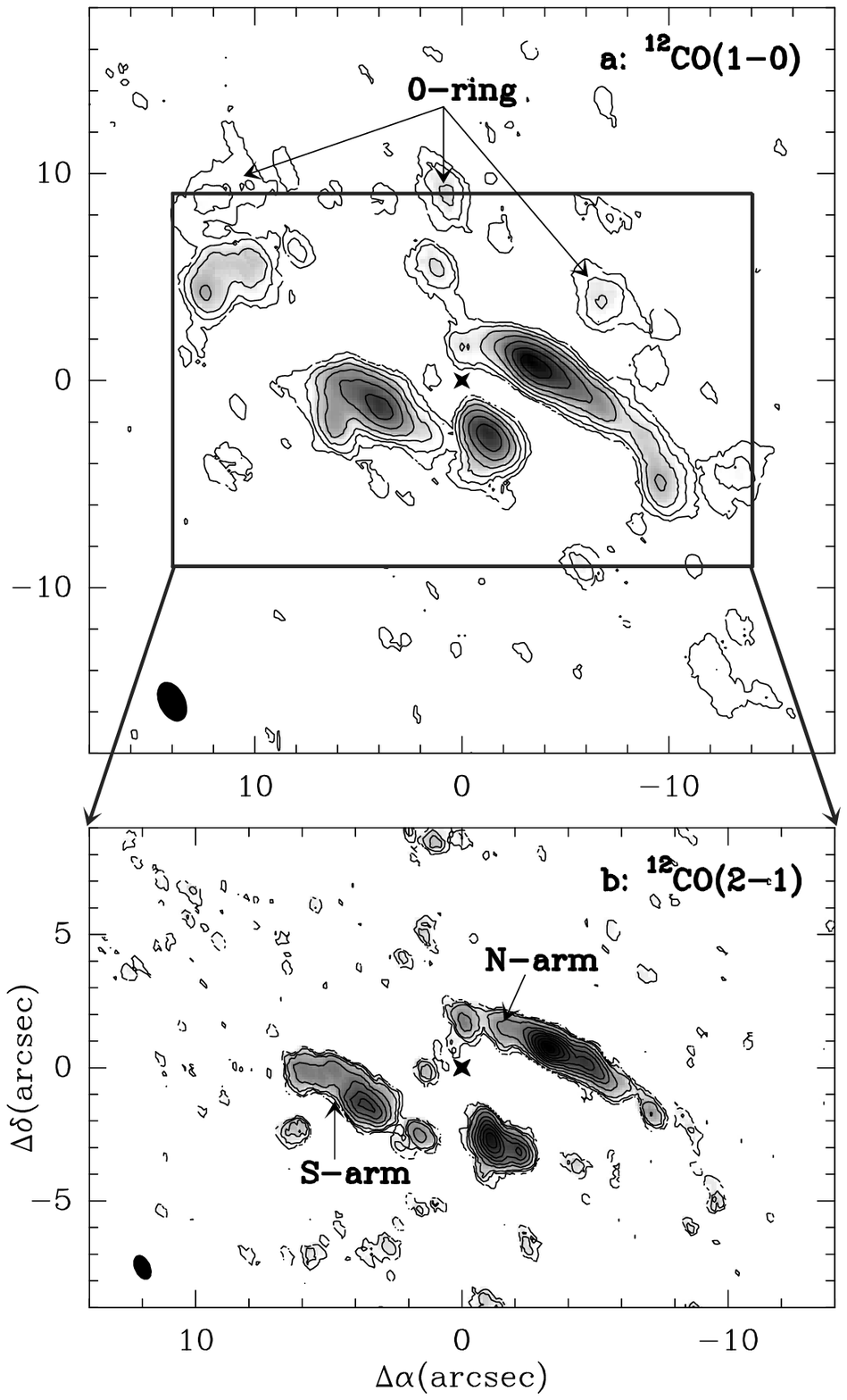}
\caption{{\bf a)}~The CO(1--0) integrated intensity map obtained with the PdBI in the nucleus of NGC\,4579. The map is shown in grey scale with contour levels 0.2, 0.5, 1, 1.7, 2.6, 3.7 and 5~Jy~km~s$^{-1}$~beam$^{-1}$.  The filled ellipse at the bottom left corner represents the CO beam size.  ($\Delta\alpha$,~$\Delta\delta$)--offsets are with respect to the location of the AGN (marked by the star): (RA$_{2000}$, Dec$_{2000}$)=(12$^{h}$37$^{m}$43.52$^{s}$, 11$^{\circ}$49$^{\prime}$05.5$\arcsec$). {\bf b)}~Same as {\bf a)} but here for the CO(2--1). Contours levels from 0.3, 0.6, 1 to 9 in steps of 1~Jy~km~s$^{-1}$~beam$^{-1}$. Positive intensity contours have been derived using a 3-$\sigma$ clipping on both data cubes. The position of the N-arm, S-arm and O-ring, as defined in the text, are highlighted.} 
%%Negative (--3.5-$\sigma$) levels have been calculated for the two data cubes with no clipping.}
        \label{fig:intensity map}
\end{figure}
%%%%%%%%%%%%%%%%%%%%%%%%%%%%%%%%%%%%%%%%%%%%%%%%%%%%%%%%%%%%%%%%%%%%%%%%%%%%%%%%%%%%%%%%%%%%%%%%%%%%%%%%%%%%%%%%%
%%%%%%%%%%%%%%%%%%%%%%%%%%%%%%%%%%%%%%%%%%%%%%%%%%%%%%%%%%%%%%%%%%%%%%%%%%%%%%%%%%%%%%%%%%%%%%%%%%%%%%%%%%%%%%%

\begin{figure}[tbh!]
\centering
    \includegraphics[width=8.5cm]{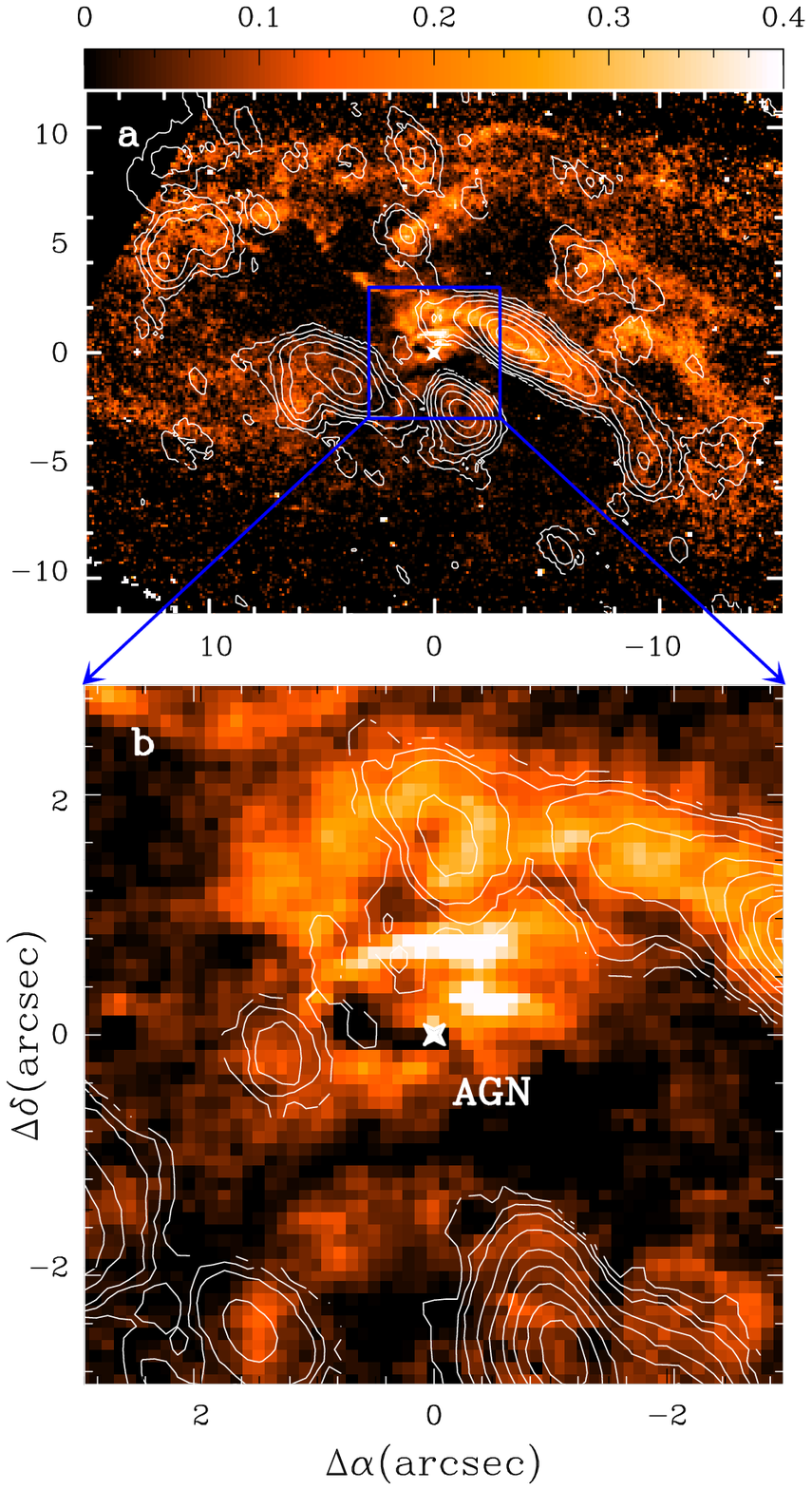}
\caption{{\bf a)}~The CO(1--0) integrated intensity map obtained with the PdBI (contour levels are 0.2, 0.5, 1.0, 1.7, 2.6, 3.7 and 5Jy~km~s$^{-1}$~beam$^{-1}$) is overlaid on the $V-I$ color image from  HST (color scale) observed in the nucleus of NGC\,4579. ($\Delta\alpha$,~$\Delta\delta$)--offsets are with respect to the location of the AGN (marked by the star). {\bf b)} Same as {\bf a)} for the CO(2--1) line with contour levels from 0.3, 0.6, 1.0 to 9 in steps of 1.0~Jy~km~s$^{-1}$~beam$^{-1}$. A close-up view of the inner 200~pc region shows a lopsided disk with the AGN lying on its southwestern edge.}
        \label{fig:overlay V-I-CO}
\end{figure}

%%%%%%%%%%%%%%%%%%%%%%%%%%%%%%%%%%%%%%%%%%%%%%%%%%%%%%%%%%%%%%%%%%%%%%%%%%%%%%%%%%%%%%%%%%%%%%%%%%%%%%%%%%%%%%%%

\subsection{IRAM-30m CO observations and short spacing correction}

With the aim of including the short spacing correction in the final image, we have mapped the emission of the 2--1 and 1--0 lines of CO in NGC\,4579 with the IRAM 30m telescope. We have covered the central 28$\arcsec$ (2.8~kpc) region of the galaxy using a grid of 5$\times$5 points with a 7$\arcsec$ spacing. The area covered is larger than the primary beam size of the PdBI at 230~GHz, and a significant fraction ($\sim$70$\%$ ) of the corresponding field of view at 115~GHz.  
Observations were carried out in two observing runs between July 2002 and June 2004. At 115~GHz and 230~GHz, the telescope half-power beam widths are 22$\arcsec$ and 12$\arcsec$, respectively.  The 3\,mm and 1\,mm receivers of the 30m telescope were tuned to the redshifted frequencies of the lines. The backends were two 1~MHz filter banks and auto-correlator spectrometers. The velocity range covered was $\rm 1340\,km\,s^{-1}$ for the 3\,mm lines and $\rm 670-1200\,km\,s^{-1}$ for the 1\,mm lines. Typical system temperatures during the observations were $\sim$250\,K at 3\,mm and $\sim$600\,K at 1\,mm. All receivers were used in single side-band mode (SSB), with a high rejection of the image band: $>$15\,dB at 1\,mm and $>$20\,dB at 3\,mm. The latter assures that the calibration accuracy for the bulk of our data is better than 20$\%$. Throughout the paper, velocity-integrated line intensities ($I$) are given in antenna temperature scale, $T_{\rm a}^{*}$. The $T_{\rm a}^{*}$ scale relates to the main beam temperature scale, $T_{\rm mb}$, by the equation $T_{\rm mb} = (F_{\rm eff}/B_{\rm eff}) T_{\rm a}^{*}$, where $F_{\rm eff}$ and $B_{\rm eff}$ are the forward and beam efficiencies of the telescope at a given frequency. For the IRAM 30m telescope $F_{\rm eff}/B_{\rm eff} = 1.27$ (1.75) at 115\,GHz (230\,GHz) and $S/T_{\rm mb} = 4.95$\,Jy\,K$^{-1}$. Wobbler switching mode was used to assure flat baselines, taking a reference position offset by 4$^{\prime}$ in azimuth. The pointing was regularly
checked on continuum sources and the accuracy was 3$\arcsec$ rms.  

Short spacings were included in the combined PdBI+30m image using the SHORT-SPACE task available in the GILDAS
software (Guilloteau \& Lucas~\cite{gui00}). The relative weights are chosen so as to guarantee that similar
absolute weights are taken for the single-dish data and the interferometer data within a ring in the UV plane going from $1.25\,D/\lambda$ to $2.5\,D/\lambda$($D$=15 m).  In the combined map we ended up recovering all the missing flux and simultaneously kept the spatial resolutions very close to those of the original PdBI maps. Throughout this paper we use the combined PdBI+30m data cube, i.e., corrected by short spacings. We estimate that the map including only the PdBI spacings, published by GB05 recovers $\sim$65$\%$ of the total CO(1--0) flux measured with the combined PdBI+30m map. A similar retrieval factor is derived for the CO(2--1) line. The 1-$\sigma$ noise levels in 10~km~s$^{-1}$-wide channels for the PdBI+30m images are 2.3\,mJy~beam$^{-1}$ in CO(1--0) and 4.6\,mJy~beam$^{-1}$ in CO(2--1).

The 30m maps probe the CO emission of molecular gas in the central $r$\,$\sim$1.4~kpc of NGC\,4579 (Fig.~\ref{fig:30m-spectra}). The emission is spatially resolved in CO(2--1) and is suggestive of an elongated gas distribution, mostly along the east-west axis. This picture is confirmed by the higher resolution interferometer data (see Sect.~\ref{CO-distribution}).

\subsection{HI maps}

The 21cm HI line emission in NGC\,4579 was observed in 2003 and 2004 using the NRAO Very Large Array (VLA) in its C and D-configurations. A first analysis of the results obtained in NGC\,4579 have been published by Haan et al.~(\cite{haa08a}) as part of a comprehensive spectroscopic imaging HI survey conducted in a sample of 16 LLAGNs including all NUGA 
galaxies. This HI survey reaches a moderate spatial resolution ($\sim$30-40'') and high sensitivity (3-$\sigma$ detection limit of $\leq$10$^{19}$\,cm$^{-2}$). Calibration procedures followed are fully described in Haan et al.~(\cite{haa08a}). The data issued from this HI survey have been recently the basis of an overall study of
the in/out-flow rates in 7 NUGA galaxies (Haan et al.~\cite{haa08b}). The HI maps of NGC\,4579 used in this paper have been produced by robustly weighted imaging of the data in order to maximize spatial resolution while keeping a high sensitivity threshold. With this procedure we reach a spatial resolution of $23.1''\times 21.6''$ and a 3-$\sigma$ sensitivity limit of 0.51~mJy beam$^{-1}$. 
Analysis was done with the Groningen Image Processing SYstem (GIPSY) as described by Haan et al.~(\cite{haa08a}). The channel maps were combined to derive an HI intensity map with a flux cut-off of 3$\sigma$. The HI emission in the disk of NGC\,4579 lies well inside the primary beam field of view of 30', therefore no attempt has been made to correct for primary beam attenuation.

\subsection{Optical, UV and IR images}

We used several optical, UV and IR images of NGC~4579 to study the star formation pattern and the stellar structure of the disk of the galaxy as listed below.  

We first acquired from the HST archive\footnote{Based on observations made with the NASA/ESA Hubble Space Telescope, 
obtained from the data archive at the Space Telescope Science Institute. 
STScI is operated by the Association of Universities for Research in 
Astronomy, Inc. under NASA contract NAS 5-26555.} broadband images of NGC\,4579, including two WFPC2 images (F547M and F791W). Similarly, a continuum-subtracted H$_{\alpha}$ emission line image was created by subtracting the associated continuum-band image, as described in Pogge et al.~(\cite{pog00}). The optical images were combined using ({\it crreject}) to eliminate cosmic rays, and calibrated according to Holtzman et al.~(\cite{hol95}). The ``pedestal'' effect 
(see B\"oker et al.~\cite{bok99}) was removed with the van der Marel algorithm\footnote{\it
http://www.stsci.edu/$_{\tilde{\,}}$marel/software/pedestal.html}. Sky values were assumed to be
zero since the galaxy filled the WFPC2 frames, an assumption which makes an error of
$\sim$\,0.1 mag at most, in the corner of the images. 

We used the ultraviolet (UV) image of NGC\,4579 obtained with the Hubble Space Telescope's Advance Camera for Surveys (ACS) and its detector High Resolution Channel (HRC) that provide $\sim$0.028$^{\prime\prime}\times0.025^{\prime\prime}$\,/pixel spatial resolution. This image correspond to the F330W band whose central wavelength is $\sim$\,3300\,~$\AA$. (Maoz et al.~\cite{mao05}). The image is available already reduced with the Space Telescope Science Institute (STScI) pipeline.

We also used the far-ultraviolet (FUV) image of the \textit{GALEX} satellite, whose band is centered at $\lambda_{eff}$ = 1516 $\AA$. This image has been obtained with a total exposure time of 1586 sec and covers a square region on the sky of size $\sim$\,$5800^{\prime\prime} \times 5800^{\prime\prime}$. The image is available already reduced with a slightly modified version of the \textit{GALEX} data pipeline (Gil de Paz et al.~\cite{gil07}), expressed in intensity units and sky-subtracted. 

In order to derive the stellar potentials, we adopted the $K$-band image obtained using the INGRID NIR camera of the William Herschel Telescope (WHT) by Knapen et al.~(\cite{kna03}). The field of view provided by the camera is 4.2$^{\prime}\times4.2^{\prime}$ with a pixel size of 0.242$^{\prime\prime}$. The details of the data reduction and calibration procedures for this image are described in Knapen et al.~(\cite{kna03}) and references therein.

IRAC images in bands 3.6 $\mu$m and 8 $\mu$m were retrieved from the Spitzer archive of NGC~4579 (Kennicutt et al.~\cite{kenn03}). Theses images cover a region of the sky of size $\sim$1400$^{\prime\prime}\times1500^{\prime\prime}$ and have pixel size of $0.75^{\prime\prime}$. Images are reduced with the SINGS IRAC pipeline and are calibrated in MJy~sr$^{-1}$ units. We have recentered on the AGN and deprojected onto the galaxy plane both images to superpose these IR images to the HI map.

\section{Dynamical center and position of the AGN}\label{AGN}

Figures ~\ref{fig:channels-CO1--0} and \ref{fig:channels-CO2--1} show the velocity-channel maps of CO(1--0) and CO(2--1) emission in the central region of NGC\,4579, respectively. The observed kinematical pattern is characteristic of a spatially resolved rotating circumnuclear disk of $\sim$2~kpc--radius. The gas kinematics indicate that the eastern (western) side of the CO disk is red (blue)-shifted with respect to the reference velocity $v_{o}$ initially assumed in this work. This is in agreement with the kinematical major axis of the circumnuclear disk being oriented roughly along the RA-axis, an orientation similar to that of the outer disk ($r$\,$>$5-10~kpc) probed by HI and H$_{\alpha}$ emission (Guhathakurta et al.~\cite{guh88}; Chemin et al.~\cite{che06}; Haan et al.~\cite{haa08a}). The global kinematics are perturbed by non-circular motions (see Sect.~\ref{CO-kinematics}). However, and contrary to previous claims (Chemin et al.~\cite{che06}), we do not find evidence that the kinematical major axis of the inner disk and that of the outer disk are substantially different. In this paper we adopt a common $PA$=95$^{\circ}$ at all radii in rough agreement with the values derived from H$_{\alpha}$ and HI (see Sect.~\ref{rotation}). 

%%%%%%%%%%%%%%%%%%%%%%%%%%%%%%%%%%%%%%%%%%%%%%%%%%%%%%%%%%%%%%%%%%%%%%%%%%%%%%%%%%%%%%%%%%%%%%%%%%%%%%%%%%%%%%%%

\begin{figure*}[thb!]
\centering
    \includegraphics[width=14.5cm]{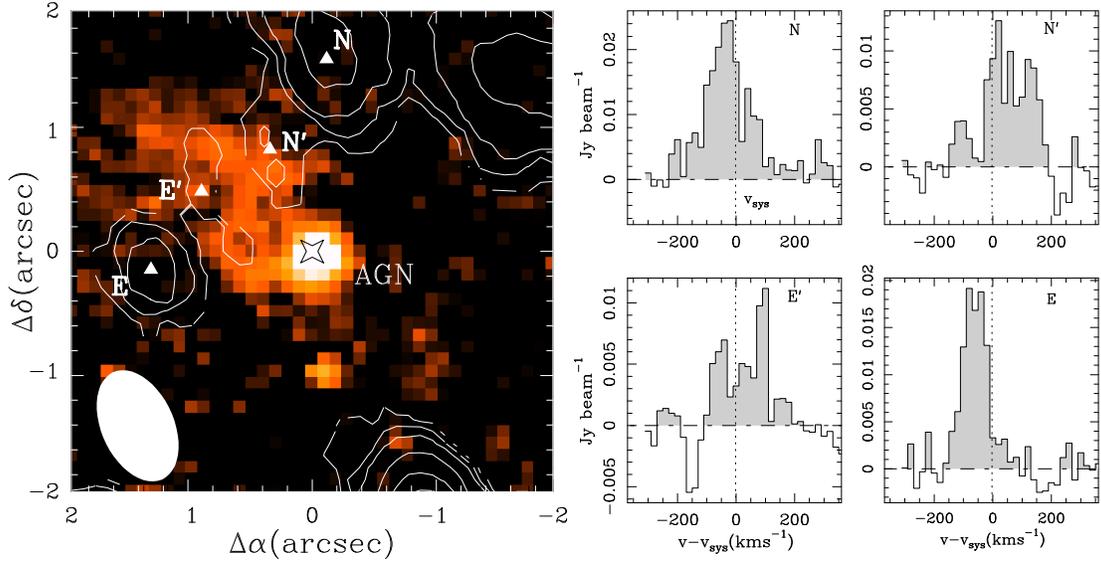}
\caption{{\bf a)}({\it Left panel})~Same contours for the CO(2--1) line as in Fig.~\ref{fig:overlay V-I-CO} overlaid on the OIII line color image from  HST of Pogge et al.~(\cite{pog00}) (color scale) observed in the nucleus of NGC\,4579. Complexes N, N', E' and E are highlighted. {\bf b)}({\it Right panels})~CO(2--1) line spectra observed towards complexes N, N', E' and E. Velocities are relative to $v_{sys}$.}
        \label{fig:overlay OIII-CO}
\end{figure*}
%%%%%%%%%%%%%%%%%%%%%%%%%%%%%%%%%%%%%%%%%%%%%%%%%%%%%%%%%%%%%%%%%%%%%%%%%%%%%%%%%%%%%%%%%%%%%%%%%%%%%%%%%%%%%%%%

The best fits for the center of symmetry of rotation and systemic velocity are ($\Delta\alpha$,$\Delta\delta$)=(--0.9$''$, 3$''$) and $v_{sys}^{LSR}$=$v_{o}$-50~km~s$^{-1}$=1470~km~s$^{-1}$, respectively. Within the errors, the derived CO dynamical center ($\alpha_{J2000}=12^{\rm h}37^{\rm m}43.52^{\rm s}$, $\delta_{J2000}=11^\circ$49$'$05.54$''$) coincides with the position of the AGN, as determined by multi-frequency very long based interferometry (VLBI) observations of the non-thermal continuum emission of NGC\,4579 (Anderson et al.~\cite{and04}; Krips et al.~\cite{kri07}). This coincidence is confirmed by our detection of a point source in continuum emission at 115\,GHz (with a flux of 11.5$\pm$0.3~mJy) and 230\,GHz (with a flux of 11.0$\pm$1~mJy) (see Fig.~\ref{fig:agn}). As fully discussed by Krips et al.~(\cite{kri07}), the spectral index of the continuum emission from 1~GHz to 230~GHz indicates that the spectrum of the AGN source is flat or slightly inverted, as frequently found in LLAGNs.

Weak but still statistically significant emission  has also been detected in the 2--1 and 1--0 lines of CO coming from a spatially unresolved source at the position of the AGN. Emission is detected at a $\geq$4-$\sigma$ level in integrated intensities over a velocity range $\Delta v\sim$150~km~s$^{-1}$. This suggests the existence of a molecular torus/disk of radius $r$\,$\leq$40\,pc, an upper limit determined by the beam size at 230\,GHz.  The systemic velocity derived from CO($v_{sys}^{LSR}$=1470$\pm$10~km~s$^{-1}$=$v_{sys}^{HEL}$=1466$\pm$10~km~s$^{-1}$) is $\sim$50~km~s$^{-1}$ blueshifted with respect to the value determined from HI ($v_{sys}^{HEL}$=1519$\pm$10~km~s$^{-1}$ from RC3). The value of $v_{sys}$ based on CO, besides symmetrizing the velocity field of the gas inside $r$\,$\sim$2~kpc, coincides with the velocity centroid of the CO(2--1) line emission detected in the torus/disk. The 1--0 spectrum towards the AGN is slightly blueshifted with respect to $v_{sys}$ due to beam smearing effects at this frequency. 
%The CO(2--1) line gives a more accurate estimate of $v_{sys}$  because of its higher spatial resolution.

\section{Molecular gas distribution}\label{CO-distribution}

The velocity-integrated CO(1--0) flux obtained in the combined PdBI+30m map within the 
42$''$ primary beam field of the PdBI is $S_{CO}$=150~Jy~km~s$^{-1}$. This is $\sim$90$\%$ of the total flux detected by Kenney \& Young~(\cite{ken89}) towards the central position of a raster map of NGC\,4579, made along the major axis of the galaxy with the Five College Radio Astronomy Observatory (FCRAO) telescope\footnote{Note that the two flux estimates are directly comparable as the FWHM of the FCRAO at 115~GHz is 45$''$, i.e., similar to the corresponding PdBI primary beam size at this frequency}. Assuming a CO-to-H$_2$ conversion factor typical of galaxy nuclei $X$=$N(H_2)/I_{CO}$=2.3$\times$10$^{20}$~cm$^{-2}$~K$^{-1}$~km$^{-1}$~s
(Solomon \& Barrett~\cite{sol91}), the total H$_{2}$ mass derived from the PdBI+30m map within the 42$''$ PdBI field of view is $M(H_2)\sim$5$\times$10$^{8}M_{\sun}$. Including the mass of helium, the corresponding total molecular gas mass is $M_{gas}$=$M(H_2+He)$=1.36$\times M(H_2)\sim$7$\times$10$^{8}M_{\sun}$.
The distribution of molecular gas is best seen in Figs.~\ref{fig:intensity map}\textit{a,b}, which show the
CO intensity maps obtained by integrating the emission in velocity
channels from $v$-$v_{sys}$\,=\,--210 to 260~km~s$^{-1}$ using a 3-$\sigma$ clipping.

The sensitivity of the CO(1--0) map is reduced beyond the primary beam radius at this frequency $r$\,=21$\arcsec$($\sim$2~kpc). This explains why we do not see in our maps the large-scale two-arm spiral structure detected in the single-dish CO(1--0) map of Kuno et al.~(\cite{kun07}). This large-scale spiral develops from $r$\,$\sim$4~kpc to $r$\,$\sim$10~kpc and shows a good correspondence with the HI pseudo-ring detected by Haan et al.~(\cite{haa08a}) (see Fig.~\ref{fig:pot_deproj}\textit{a}). Normalized to a distance $D$=20~Mpc, we estimate that the total molecular gas mass inside $r$\,$\sim$10~kpc amounts to $M_{gas}\sim$4$\times$10$^{9}M_{\sun}$.

Inside $r$\,$\sim$2~kpc we identify in the PdBI maps three main components in the distribution of the 2--1 and 1--0 line emission:  
\begin{itemize}
\item

{\it The central disk:}~the 2--1 line map of Fig.~\ref{fig:intensity map}\textit{b}  gives a sharp view of the distribution of molecular gas at radii $r$\,$<$\,200~pc. 
As shown in Figs.~\ref{fig:overlay V-I-CO} and \ref{fig:overlay OIII-CO}, there is little molecular gas at $r$\,$<$200~pc from the central engine. The most massive gas complexes, of a few$\sim$10$^{6}M_{\odot}$, lie east and north of the nucleus at $r$\,$\sim$150\,pc (complexes E and N in Fig.\ref{fig:overlay OIII-CO}, respectively). Complexes E and N 
are connected by an arc of low-level CO(2-1) emission (complexes E' and N' in Fig.~\ref{fig:overlay OIII-CO}). These constitute altogether the molecular gas counterpart of a lopsided disk that can be clearly identified as a dusty feature in the V--I color HST image of the galaxy (Fig.~\ref{fig:overlay V-I-CO}). The AGN lies close to the southwestern edge of the central disk.  As discussed in Sect.~\ref{AGN}, there is also weak CO emission related to the AGN; this corresponds to a molecular gas torus or disk of $M_{gas}$=$M(H_2+He)\sim$10$^{6}M_{\sun}$.

\item

{\it The inner spiral arms:}~the bulk ($>$80$\%$) of the molecular gas as traced by the two CO lines is piled up in two highly contrasted spiral lanes that lie at the leading edges of the $\sim$12~kpc-diameter bar, identified in the $K$-band image of Knapen et al.~(\cite{kna03})(see also Eskridge et al.~\cite{esk02} and Jarrett et al.~\cite{jar03}).
 The described geometry is also illustrated in Fig.~\ref{fig:pot_deproj}\textit{b,c}, which shows the deprojected version of the CO maps overlaid on the $K$-band image of the galaxy. The CO arms develop from $r$\,$\sim$300~pc to $r$\,$\sim$1.3~kpc. At closer examination, it can be seen that the CO spiral lane stretching north of the AGN (hereafter denoted as N-arm) is more regular and better delineated than its southern counterpart (hereafter referred to as S-arm). Moreover, the inner segment of the N-arm is significantly closer to the AGN compared to that of the S-arm (Fig.~\ref{fig:overlay V-I-CO}). As discussed in Sect.~\ref{tracers} (see also Fig.~\ref{fig:overlay Ha-CO}), a similar asymmetry is observed in H$_{\alpha}$ emission. Furthermore, the N-arm is well correlated with a northern dust lane characterized by a red V--I color, as derived from the HST images of the galaxy (Fig.~\ref{fig:overlay V-I-CO}). In contrast, the S-arm can be hardly identified in the V--I color image of Fig.~\ref{fig:overlay V-I-CO}. This overall north-south asymmetry in the color image can be partly explained if we assume that the northern side of the disk is the near side of the galaxy. However, the differences in the morphology of the N-arm and the S-arm, revealed by CO and H$_{\alpha}$, are to a large extent intrinsic and thus cannot be explained by a projection effect.

%%%%%%%%%%%%%%%%%%%%%%%%%%%%%%%%%%%%%%%%%%%%%%%%%%%%%%%%%%%%%%%%%%%%%%%%%%%%%%%%%%%%%%%%%%%%%%%%%%%%%%%%%%%%%%%%

\begin{figure}[tbh!]
\centering
    \includegraphics[width=8.5cm]{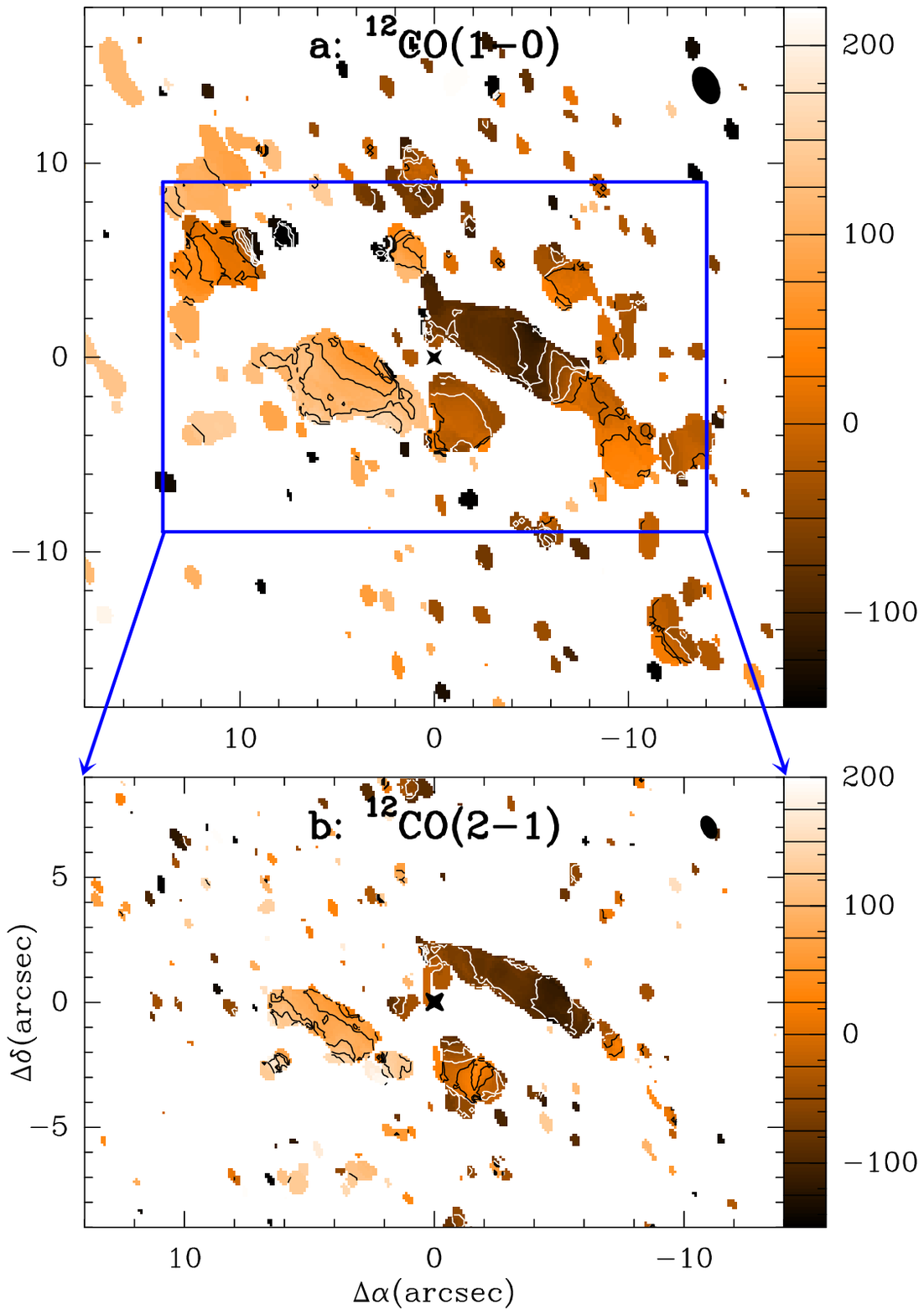}
\caption{
	{\bf a)}~CO(1--0) ({\it Upper panel}) and CO(2--1) ({\it Lower panel}) isovelocities contoured over false-color velocity maps. Velocities span the range (--225~km~s$^{-1}$, 225~km~s$^{-1}$) in steps of 25~km~s$^{-1}$. Velocity scale is relative to $v_{sys}$. Isovelocities have been derived using a 4-$\sigma$ clipping on both data cubes. The AGN position is marked with a star.}
        \label{fig:isovelocity map}
\end{figure}
%%%%%%%%%%%%%%%%%%%%%%%%%%%%%%%%%%%%%%%%%%%%%%%%%%%%%%%%%%%%%%%%%%%%%%%%%%%%%%%%%%%%%%%%%%%%%%%%%%%%%%%%%%%%%%%%

%%%%%%%%%%%%%%%%%%%%%%%%%%%%%%%%%%%%%%%%%%%%%%%%%%%%%%%%%%%%%%%%%%%%%%%%%%%%%%%%%%%%%%%%%%%%%%%%%%%%%%%%%%%%%%%%
\begin{figure*}[tbh!]
\centering
    \includegraphics[width=16.5cm]{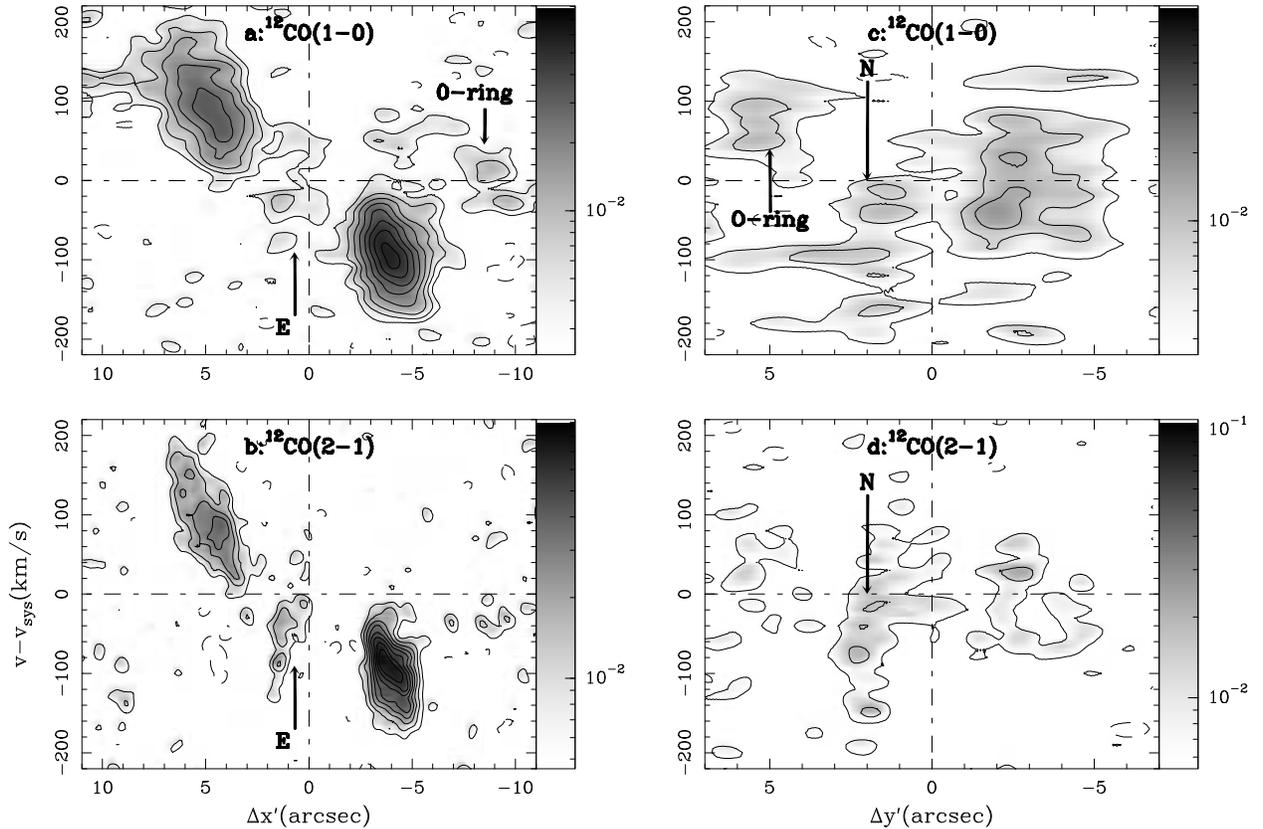}
\caption{{\bf a)}~Position-velocity diagram of CO(1--0) along the mayor axis of NGC~4579 ($PA$=95$^{\circ}$). Contour levels go from --1.5$\sigma$, 1.5$\sigma$, 3$\sigma$, 5$\sigma$, 8$\sigma$ to 24$\sigma$ in steps of 4$\sigma$ where the 1-sigma rms $\sigma$=2.3~mJy~beam$^{-1}$. {\bf b)} Same as {\bf a)} but for CO(2--1) line with levels going from --1.5$\sigma$, 1.5$\sigma$, 3$\sigma$, 5$\sigma$, 7$\sigma$ to 15$\sigma$ in steps of 2$\sigma$ where the 1-sigma rms $\sigma$=4.6~mJy~beam$^{-1}$. 
{\bf c)}~Position-velocity diagram of CO(1--0) along the minor axis of NGC~4579 ($PA$=5$^{\circ}$). 
Contour levels are --1.5$\sigma$, 1.5$\sigma$, 3$\sigma$ and 5$\sigma$.
{\bf d)} Same as {\bf c)} but for the CO(2--1) line. Contour levels are --1.5$\sigma$, 1.5$\sigma$ and 3$\sigma$.
In all panels velocities are relative to the systemic velocity ($v_{sys}^{LSR}$=1470~km~s$^{-1}$) and both $\Delta$x$^{\prime}$ and $\Delta$y$^{\prime}$ offsets along the major and minor axis, respectively, are relative to the AGN locus. The position of the N and E complexes and that of the O-ring are highlighted.}
        \label{fig:kinematics p-v}
   \end{figure*}
%%%%%%%%%%%%%%%%%%%%%%%%%%%%%%%%%%%%%%%%%%%%%%%%%%%%%%%%%%%%%%%%%%%%%%%%%%%%%%%%%%%%%%%%%%%%%%%%%%%%%%%%%%%%%%%%

\item

{\it The outer ring:}~further out, north, there is a chain of emission clumps detected in CO(1-0) (hereafter referred to as outer ring or O-ring). This component, only detected in the 1--0 line, has no clear southern equivalent.  As shown in  Fig.~\ref{fig:overlay V-I-CO}, the O-ring is well correlated with a chain of gas spurs characterized by a red V--I color. Furthermore, the O-ring is also detected in H$_{\alpha}$ emission (see  Fig.~\ref{fig:overlay Ha-CO} and discussion in Sect.~\ref{tracers}); this morphology, first described by Pogge~(\cite{pog89}) as {\it unusual}, mimics a one-sided ellipsoidal ring.

\end{itemize}

In summary, the bulk of the gas response in the disk of NGC\,4579, as traced by the PdBI CO maps described above, follows the expected gas flow pattern induced by the bar potential of the galaxy. In particular, the inner spiral arms appear as two offset gas lanes which are leading with respect to the bar. This geometry suggests that there is spatially extended ILR region in the nucleus, i.e., leaving room for two ILRs. In terms of bar orbit structure, this result can be interpreted as an indication that there are stable stellar orbits whose major axis is perpendicular to the bar major axis ($x_2$ orbits)(e.g., Athanassoula~\cite{ath92}; Buta \& Combes~\cite{but96}; Regan \& Teuben~\cite{reg03}). However, we have identified in the CO maps departures from the purely $m$=2 point-symmetric gas flow pattern dictated by the bar. The reported deviations suggest that a superposed lopsided ($m$=1) instability may be at work. Firstly, there are indications of lopsidedness in the inner CO spiral arms, echoed by other tracers of the ISM. The O-ring is an asymmetric pattern that could be the relic of a bar resonance, which is currently being depopulated. 
%or, alternatively, the remnants of infalling material onto the disk after a minor merger episode. 
Closer to the AGN, the overall morphology of the central disk also suggests that the $m$=2 point-symmetry of the gas flow breaks up at $r$\,$<$200\,pc letting lopsidedness take over.

\section{CO line ratios}\label{CO-ratios}

The value of the 2--1/1--0 integrated intensity ratio ($R_{21}$) is obtained after convolving the 2--1 map to the lower resolution of the 1--0  map, both maps including short spacings. $R_{21}$ ranges from $\sim$0.4 to $\sim$1 inside the observed region (in $T_{mb}$~km~s$^{-1}$ units); the average ratio is $\sim$0.6. This ratio is close to the canonical value observed in spiral disks and is typical of optically thick molecular gas clouds (e.g., Braine \& Combes~\cite{bra92}; Garc\'{\i}a-Burillo et al.~\cite{gb03}). There is a radial trend in the excitation of the gas, as measured by the line ratio: close to the AGN, at $r$\,$<$200~pc, $R_{21}$\,$\sim$0.8--1 (over the central disk), whereas $R_{21}$\,$\sim$0.4--0.6 at $r$\,$>$500~pc (over the S-arm, the N-arm and the O-ring). Hints of a similar trend are found in other NUGA targets (e.g., Garc\'{\i}a-Burillo et al.~\cite{gb03b}) and can be interpreted as evidence for external heating of molecular clouds by X-rays in the vicinity of the AGN (Baker et al.~\cite{bak03}).

%An independent confirmation of this picture comes from the new HST image of the nuclear region of NGC\,4579 
%obtained with the ACS camera at 3300 \AA; this image resolves the central disk into a winding $m=1$ spiral 
%instability that mimics a ring (Contini \cite{con04}).

\section{Molecular gas kinematics}\label{CO-kinematics}

\subsection{Rotation curve and dynamical mass}\label{rotation}

We show in Fig.~\ref{fig:isovelocity map} the isovelocity contour maps obtained from the CO(1--0) 
and CO(2--1) data in the nucleus of NGC\,4579 (applying a 4-$\sigma$ clipping). 
The kinematics of the molecular gas are consistent with those of a spatially resolved 
rotating disk. The sense of the rotation of the gas in the galaxy plane
is counterclockwise assuming the northern side is the near side (see Sect.~\ref{CO-distribution}).

\begin{figure}[tbh!]
\centering
    \includegraphics[width=8.5cm]{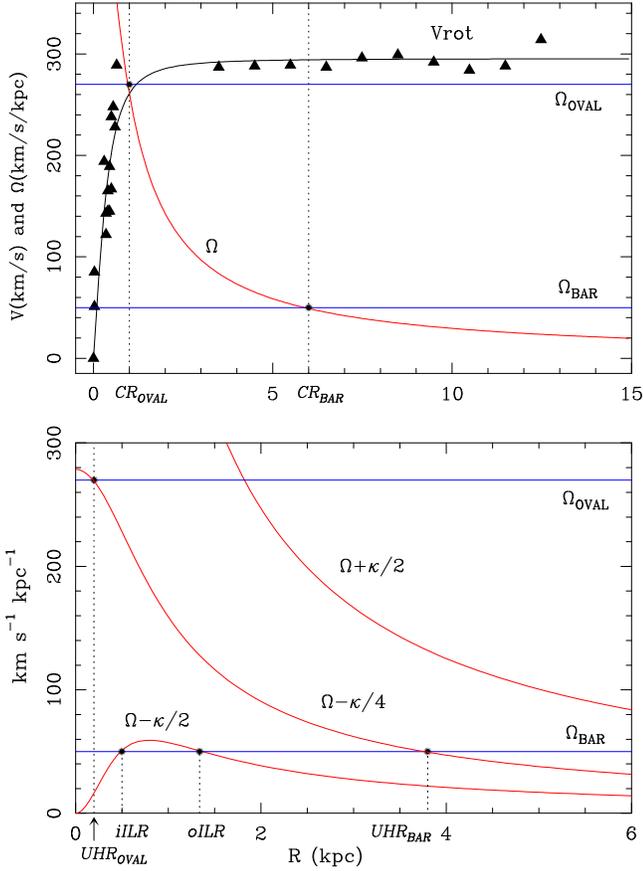}
\caption{{\bf a)}({\it Upper panel})~Rotation curve $v_{rot}$ and angular speed curve $\Omega$ fitted from the CO and HI data of NGC~4579 (triangles are data points). A value for the bar pattern speed $\Omega_{BAR}\sim$50~km~s$^{-1}$~kpc$^{-1}$ puts the corotation radius of the bar at $r_{CR}\sim$6~kpc. Similarly, 
a corresponding pattern speed value of $\Omega_{OVAL}\sim$270~km~s$^{-1}$~kpc$^{-1}$ puts the corotation radius of the oval at $\sim$1~kpc. {\bf b)}({\it Lower panel})~Frequency curves $\Omega\pm\kappa/2$ and $\Omega-\kappa/4$.  If $\Omega_{BAR}\sim$50~km~s$^{-1}$~kpc$^{-1}$, the Ultra Harmonic Resonance (UHR) of the bar lies at $r_{UHR}\sim$3.8~kpc. We estimate the existence of an extended ILR region with 
two ILRs lying at $r_{iILR}\sim$500~pc (inner ILR=iILR) and $r_{oILR}\sim$1.3~kpc (outer ILR=oILR), respectively. With $\Omega_{OVAL}\sim$270~km~s$^{-1}$~kpc$^{-1}$ the UHR of the oval is at $\sim$200~pc.}
        \label{fig:frequencies}
   \end{figure}

An estimate of the CO rotation curve ($v_{rot}$) can be obtained from position-velocity ($p-v$) diagrams taken along the major axis of NGC\,4579 (Fig.~\ref{fig:kinematics p-v}), assuming that circular motions dominate the gas kinematics. We have fitted the major axis position angle as
$PA$=95$\pm$10$^{\circ}$; this value maximizes the line-of-sight velocity gradient within $r$\,=2~kpc. Furthermore, this determination is consistent within the errors with previous findings based on H$_{\alpha}$ and HI kinematics (Guhathakurta et al.~\cite{guh88}; Chemin et al.~\cite{che06}; Haan et al.~\cite{haa08a}). Terminal velocities ($v_{term}$) are derived by fitting multiple Gaussian profiles to the spectra across the CO major axis and selecting at each position the best estimate for $v_{term}$, taking into account velocity dispersion. During the fitting process we have masked the CO spectra between 0.5$\arcsec$(50~pc) and 2.5$\arcsec$(250~pc) east of the AGN because strong non-circular motions in this region make the gas apparently 
counter-rotate around the nucleus (see Fig.~\ref{fig:kinematics p-v} and discussion in Sect.~\ref{streaming}).
For the same reasons, the data points from the O-ring have not been included in the fit.
The velocity centroids, corrected for inclination $i$=36$^{\circ}$, and referred to $v$=$v_{sys}$, give an estimate of $v_{rot}$ for each offset along the major axis. A simple rotation curve model can be fitted using the CO data from both sides of the major axis. We choose a rotation law characteristic of a logarithmic gravitational potential, described by:

\begin{equation}
v_{rot}(r)= \frac{v_{max}r}{\sqrt{r^2+r_{max}^2}}
\end{equation}

%\begin{equation}
%v_{rot}(r)=\frac{r \times v_{max}}{r_{max}} \left[ 1/3+2/3 \biggl(\frac{r}{r_{max}}\biggr)^n %\right]^{-3/2n} 
%\end{equation} 

where $v_{max}$ is the maximum rotation velocity and $r_{max}$ is a radial scale whose value controls the steepness of $v_{rot}$. To extend $v_{rot}$ to the outer radii of the galaxy we have included in the global fit the HI data points of Guhathakurta et al.~(\cite{guh88}) and Haan et al.~(\cite{haa08a}). The best fit gives $v_{max}$=295~km~s$^{-1}$ and  $r_{max}$=530~pc (Fig.~\ref{fig:frequencies}).
The $v_{rot}$ curve shows a gradual increase to $r$\,$\sim$0.65~kpc, where $v_{rot}$\,$\sim$240~km~s$^{-1}$.
%%%%%There are few data points constraining $v_{rot}$ from $r$\,$\sim$0.65~kpc to $r$\,$\sim$3~kpc. 
From $r$\,$\sim$3~kpc to the outer disk traced by HI, up to $r$\,$\sim$12~kpc, $v_{rot}$ remains roughly flat at $\sim$295~km~s$^{-1}$. The mass inside a certain radius $r$ can be inferred using $M(r)$=$C\times r \times v_{rot}^2$/$G$, where $G$ is the constant of gravity, $M(r)$ is the mass inside a sphere of radius $r$, and $C$ is a constant varying between 0.6 and 1, depending on the disk mass model assumed. If we take a value of $C$=0.8 intermediate between the values
appropriate for spherical (1) and flat disk (0.6) distributions, $M$($r$\,=2\,~kpc)$\sim$4$\times$10$^{10}M_{\sun}$. 
This implies that the molecular gas mass fraction inside $r$\,$=$2~kpc is $\sim$2$\%$. This fraction decreases to
$\leq$0.6$\%$ at $r$\,=50~pc. 

%%%%%%%%%%%%%%%%%%%%%%%%%
%%%%%%%%%%%%%%%%%%%%%%%%%%%%%%%%%%%%%%%%%%%%%%%%%%%%%%%%%%%%%%%%%%%%%%%%%%%%%%%%%%%%%%%%%%%%%%%%%%%%%%%%%%%%%%%%

%%%%%%%%%%%%%%%%%%%%%%%%%%%%%%%%%%%%%%%%%%%%%%%%%%%%%%%%%%%%%%%%%%%%%%%%%%%%%%%%%%%%%%%%%%%%%%%%%%%%%%%%%%%%%%%%

%%%%%%%%%%%%%%%%%%%%%%%%%%%%%%%%%%%%%%%%%%%%%%%%%%%%%%%%%%%%%%%%%%%%%%%%%%%%%%%%%%%%%%
\begin{figure}[t!]
\centering
   \includegraphics[width=7.5cm]{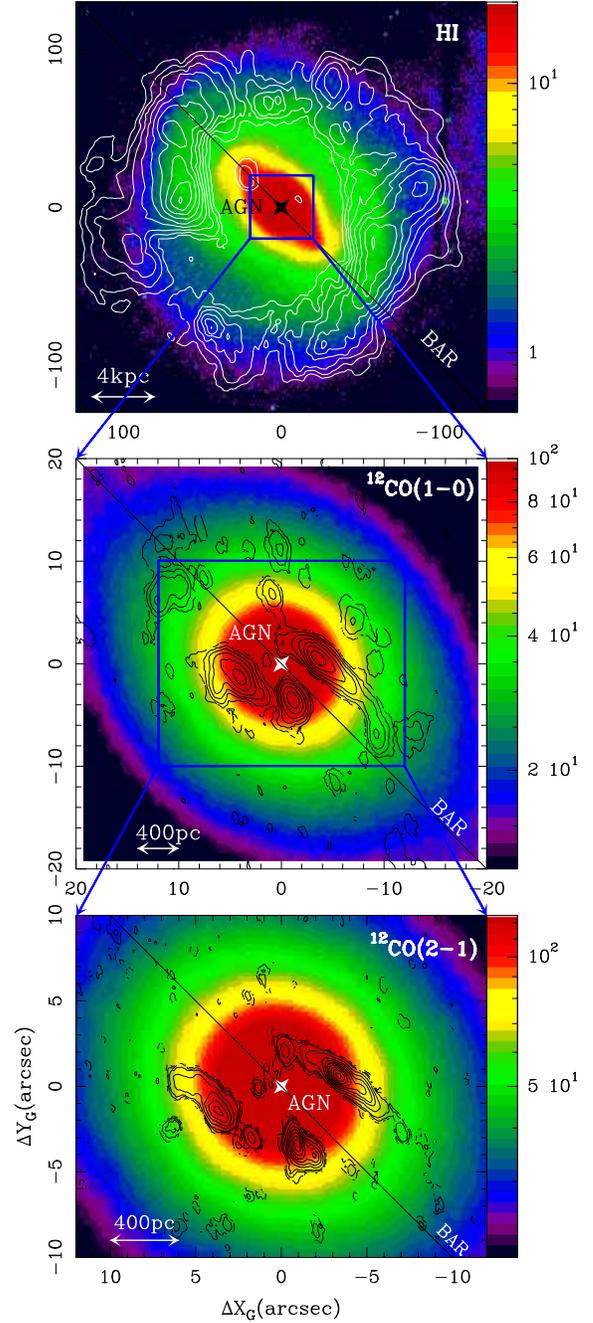}
\caption{{\bf a)} We overlay the HI intensity map obtained with the VLA (in contours from 0.04 to 0.28 in steps of 0.03~Jy~km~s$^{-1}$~beam$^{-1}$) on the $K$-band WHT image (color scale) obtained for the nucleus of NGC\,4579; both images have been deprojected onto the galaxy plane. Units on X/Y axes ($\Delta$X$_G$/$\Delta$Y$_G$) correspond to arcsec offsets along the major/minor axes with respect to the AGN. The orientation of the large-scale bar--BAR--is shown. {\bf b)} Same as {\bf a)} but we overlaid the CO(1--0) PdBI intensity map (contour levels=0.2, 0.5, 1, 1.7, 2.6, 3.7 and 5~Jy~km~s$^{-1}$~beam$^{-1}$). {\bf c)} Same as {\bf a)} but for the CO(2--1) PdBI intensity map (contour levels=0.3, 0.6, 1 to 9 in steps of 1~Jy~km~s$^{-1}$~beam$^{-1}$).}
        \label{fig:pot_deproj}
\end{figure}

%%%%%%%%%%%%%%%%%%%%%%%%%%%%%%%%%%%%%%%%%%%%%%%%%%%%%%%%%%%%%%%%%%%%%%%%%%%%%%%%%%%%%%%%%

 From the best fit of $v_{rot}$, we have derived the frequency curves, shown in Fig.~\ref{fig:frequencies}: $\Omega$, $\Omega\pm\kappa/2$ and $\Omega-\kappa/4$. These curves can be used to estimate the position of the main resonances in the disk in the context of the epicyclic approximation as discussed in Sect.~\ref{tracers}, where we compare the morphology of the molecular disk to the distribution shown by other tracers of the ISM and the stellar structure in the disk of NGC\,4579. 

%The same frequency curves, and the inferred resonances, can be obtained from an axisymmetric mass model based on %the $K$-band image of the galaxy (Sect.~\ref{grav}). However, the resonance loci are similar to those derived %directly from the $v_{rot}$ curve, and in the following we will use the latter.

\subsection{Streaming motions}\label{streaming}

The kinematics of molecular gas in the central $r$\,$\sim$2~kpc of NGC\,4579 reveal the presence of 
non-circular motions that modulate the overall rotation pattern of the disk described in Sect.~\ref{rotation}. Deviations are detected with uneven strength over the inner spiral arms (N-arm and S-arm), the outer arm (O-ring), and, most noticeably, over the central disk, as described below:

\begin{itemize}
\item
{\it The central disk:}~the radial velocities of the gas in the central disk deviate by up to $>$100kms$^{-1}$ from the expected pattern of circular rotation. CO emission from the E complex, mostly prominent in the 2--1 line, is detected at negative radial velocities ($\sim$--70~kms$^{-1}$) that are {\it forbidden} by circular rotation at this location of the disk (see Figs.~\ref{fig:overlay OIII-CO} and ~\ref{fig:kinematics p-v}). If the gas flows inside the plane, this would imply that the gas at E is counter-rotating at a deprojected speed of $v$\,$\sim$120kms$^{-1}$. Complex N, which lies close to the kinematic minor axis of the galaxy, also shows non-circular motions: instead of a zero average velocity, we measured $\sim$--50~kms$^{-1}$ (see Figs.~\ref{fig:overlay OIII-CO} and ~\ref{fig:kinematics p-v}); this can be interpreted as gas moving outward within the plane.  The CO kinematics observed in complexes N' and E' indicate instead positive radial velocities for the gas: $\sim$+75~kms$^{-1}$ at N' and $\sim$+25~kms$^{-1}$ at E'. Altogether, the radial velocities measured along a line intersecting complexes N, N', E' and E show a gradient that is reversed with respect to that expected if circular rotation prevailed.

\item
{\it The inner spiral arms:}~the isovelocity maps of Fig.~\ref{fig:isovelocity map} reveal a steep transverse velocity gradient across the S-arm. An inspection of the major axis p--v diagrams of Fig.~\ref{fig:kinematics p-v}\textit{a,b} indicates that the gas is strongly slowed down to zero rotation velocities when it encounters the S-~arm at $\Delta$x'$\sim$+3$\arcsec$. Downstream the S-arm (at $\Delta$x'$\sim$+7$\arcsec$), gas is accelerated: rotation velocities $v_{rot}$=($v~-~v_{sys}$)/sin($i$)$\sim$170/sin(36)$\sim$290~km~s$^{-1}$ are reached when the gas leaves the arm. A similar but more shallow transverse gradient is detected across the N-arm. 
\item

{\it The outer ring:}~Figure~\ref{fig:isovelocity map} shows that the O-ring is also associated with peculiar motions. The radial velocities measured in adjacent regions of the N-arm and O-ring can differ by up to 150~km~s$^{-1}$. This is illustrated in Fig.~\ref{fig:kinematics p-v} which shows the p-v plots along the major and minor axis of the galaxy.

\end{itemize}

The streaming motions detected over the inner spiral arms are in agreement with the expected kinematic signatures of the stellar bar on the gas flow within the ILR region, and thus, well inside corotation of the bar (e.g., see Canzian~\cite{can93}). However, the kinematic differences between the N-arm and the S-arm point to the influence of a superposed lopsided instability that could also be at play in the central disk component. The kinematic signature of lopsidedness is also found in the O-ring. Non-circular motions let the O-ring appear as a separate component decoupled from the inner gas response (N-arm, S-arm). 

%By themselves, the reported deviations are reminiscent of an external origin for the gas in this asymmetric 
%feature (Sect.~\ref{tracers})

%%%%%%%%%%%%%%%%%%%%%%%%%%%%%%%%%%%%%%%%%%%%%%%%%%%%%%%%%%%%%%%%%%%%%%

\begin{figure}[tbh!]
\centering
     \includegraphics[width=8.5cm]{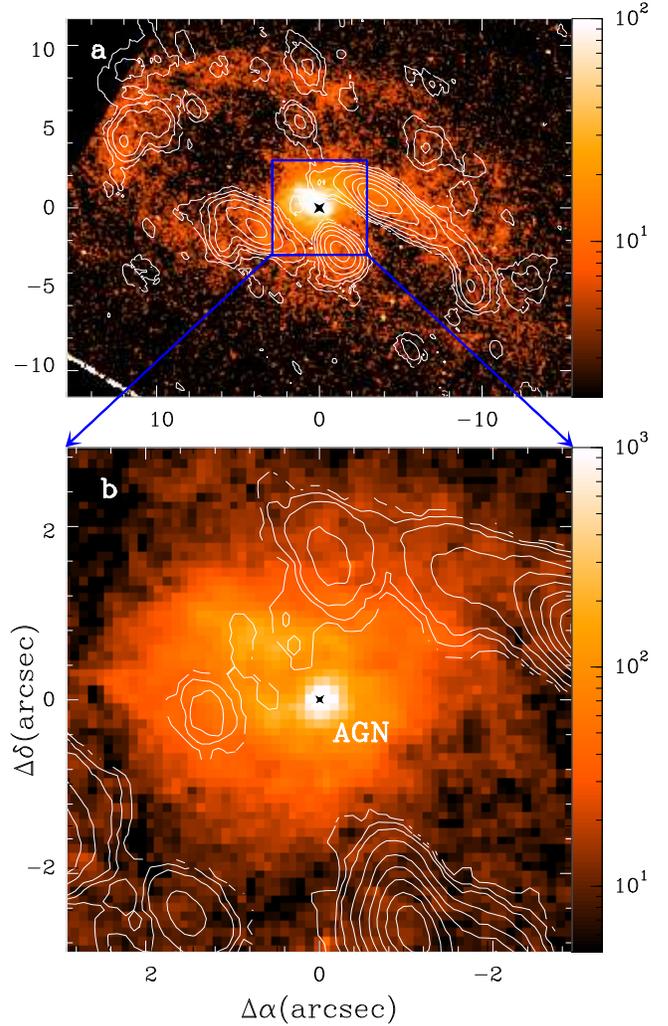}
\caption{Same contours as in Fig.~\ref{fig:overlay V-I-CO} overlaid on the H$_{\alpha}$ image from  HST (color scale) observed in the nucleus of NGC\,4579.}
        \label{fig:overlay Ha-CO}
\end{figure}

%%%%%%%%%%%%%%%%%%%%%%%%%%%%%%%%%%%%%%%%%%%%%%%%%%%%%%%%%%%%%%%%%

%%%%%%%%%%%%%%%%%%%%%%%%%%%%%%%%%%%%%%%%%%%%%%%%%%%%%%%%%%%%%%%%%%%%%%%%%%%%%%%%%%%%%%%%%
\begin{figure}[tbh!]
\centering
    \includegraphics[width=8.5cm]{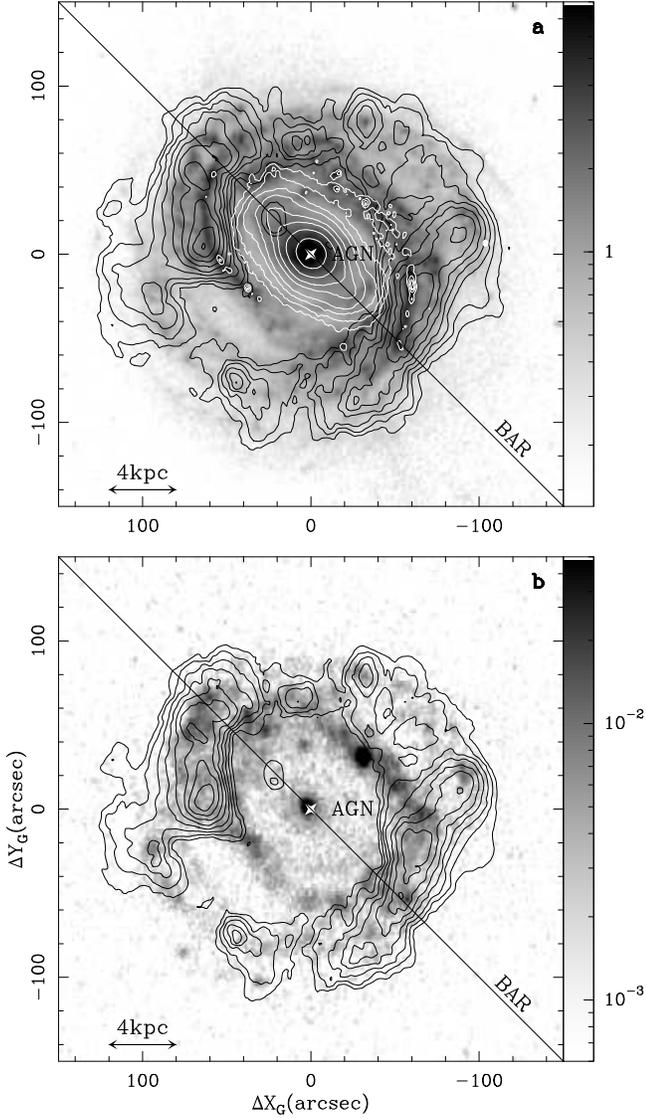} 
%{n4579-HI-Spitzer-IRAC8-3.6mic-grey.eps}
\caption{{\bf a)}({\it Upper panel})~Superposition of the HI intensity map (black contours), the 3.6 $\mu$m \textit{Spitzer/IRAC} image in contour levels 1.1, 1.3, 1.6, 2, 3, 4, 6 and 11~MJy/sr (white contours) and the 8 $\mu$m \textit{Spitzer/IRAC} image (grey scale). All images have been deprojected onto the galaxy plane.{\bf b)}({\it Lower panel})~Same as {\bf a)} but HI intensity map (in contours) superposed on the FUV GALEX image (grey scale).}
        \label{fig:overlay Spitzer/IRAC-HI}
\end{figure}
%%%%%%%%%%%%%%%%%%%%%%%%%%%%%%%%%%%%%%%%%%%%%%%%%%%%%%%%%%%%%%%%%%%%%%%%%%%%%%%%%%%%%%%%%

%%%%%%%%%%%%%%%%%%%%%%%%%%%%%%%%%%%%%%%%%%%%%%%%%%%%%%%%%%%%%%%%%%%%
\begin{figure}[tbh!]
\centering
    \includegraphics[width=8.5cm]{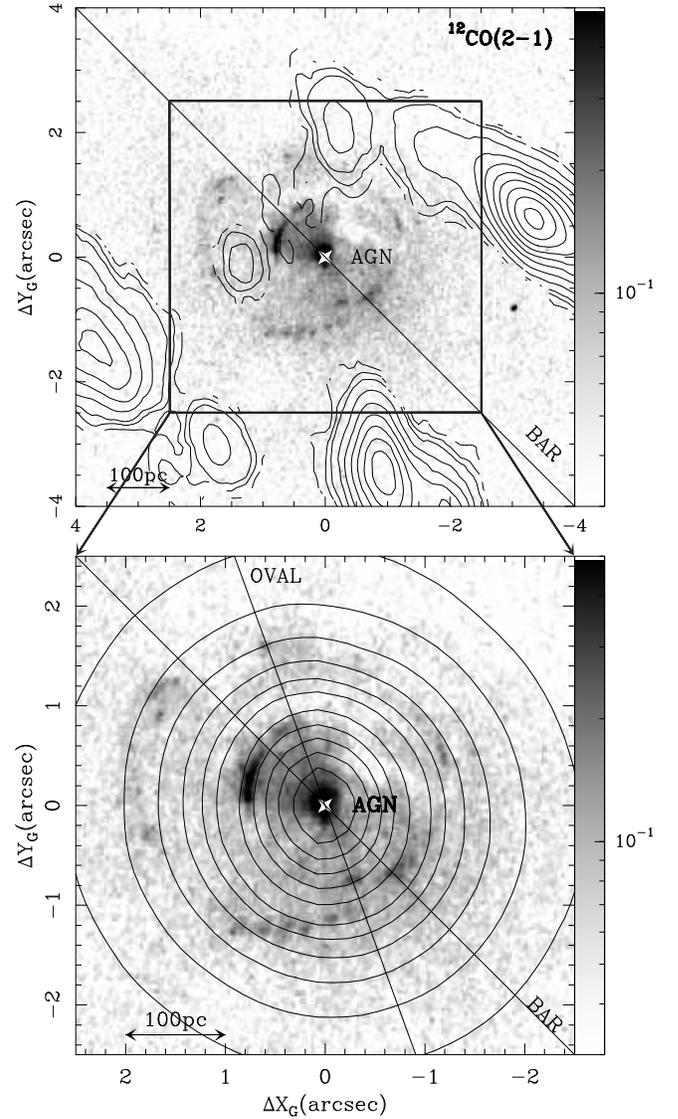}
\caption{{\bf a)}({\it Upper panel})~The $^{12}$CO(2--1) integrated intensity map obtained with the PdBI (same contours as in Fig.~\ref{fig:overlay V-I-CO}) is overlaid on the UV image of the central region of NGC\,4579 obtained with the ACS/HRC in the F330W band (in grey scale). {\bf b)}({\it Lower panel})~The $K$-band contours are overlaid on the UV-image of the galaxy. We highlight the orientation of the bar and the oval. All images have been deprojected onto the galaxy plane.}
        \label{fig:overlay UV-CO21}
\end{figure}
%%%%%%%%%%%%%%%%%%%%%%%%%%%%%%%%%%%%%%%%%%%%%%%%%%%%%%%%%%%%%%%%%%

\subsection{An outflow of molecular gas in NGC~4579?}\label{ouflow}

Alternatively, the overall morphology and velocity pattern of the central disk could be qualitatively explained by a scenario where the gas is flowing away from the nucleus in extra-planar motions, as mentioned by GB05. The OIII line and UV continuum maps obtained with HST can be used to tentatively identify a one-sided fan-shaped morphology in the emission coming from the central $r$\,$\sim$100~pc of NGC\,4579 (Pogge et al.~\cite{pog00}; Contini~\cite{con04}), suggestive of an AGN-powered outflow of hot gas. The morphology of the neutral gas disk at $r$\,$\leq$150~pc, evidenced by the $V-I$ color image of the galaxy, would indicate that the gas is piling up in a shell. The shell could be a signature of gas entrainment by the outflow (see Fig.~\ref{fig:overlay OIII-CO}). The expanding shell scenario can explain the kinematics of the N and E complexes as a signature of out-of-the-plane motions projected along the line of sight towards the observer. The positive radial velocities measured towards complexes N' and E' can be explained if out of the plane motions are projected away from the observer. In this scenario AGN feeding would be presently quenched on these scales.

The outflow scenario for NGC\,4579 is not flawless, however. Firstly, based on the current data, the putative outflow of hot gas in NGC\,4579 does not have the mass and energies typically required to produce a corresponding outflow in the neutral gas (e.g., see the case of NGC\,4528 discussed by Krause et al.~\cite{kra07}). Furthermore, in NGC\,4579 there is no evidence of a radio jet, which should be the driving agent of the outflow. Finally, there is no information on the kinematics of the ionized gas that may be indicative of outflow motions. On the whole, the outflow scenario can be considered as unlikely and merely speculative.

\section{Comparison with other tracers}\label{tracers}

In this section we compare the CO maps of NGC\,4579 with the distribution shown by other tracers of the ISM and with the stellar structure of the galaxy. This information is used to estimate the location of the resonances in NGC\,4579, a key to the interpretation of the gas-flow in the disk of the galaxy.

\subsection{The large-scale bar}\label{tracers-bar}

The 2.2$\mu$m $K$-band image of NGC\,4579 published by Knapen et al.~\cite{kna03} shows a prominent stellar bar in the disk. The bar is oriented along $PA_{bar}$\,$\sim$56$\pm5^{\circ}$ (measured from north in the plane of the sky). Once deprojected onto the plane of the galaxy (assuming $PA$=95$^{\circ}$ and $i$=36$^{\circ}$), the position angle of the bar ($PA'_{bar}$) measured from the major axis becomes: tan($PA'_{bar}$)=tan($PA$-$PA_{bar}$)/cos($i$), which implies $PA'_{bar}$\,$\sim$45$^{\circ}$, as shown in Fig.~\ref{fig:pot_deproj}. An isophotal analysis made on the $K$-band image of the galaxy gives an approximate diameter-size for the bar of $D\sim$10--12~kpc. This coincides with the value derived from a similar analysis made on the 3.6$\mu$m Spitzer/IRAC image (Fig.~\ref{fig:overlay Spitzer/IRAC-HI}). This bar size is compatible within the errors with the value calculated from the Fourier decomposition of the potential, which favors a slightly larger bar ($D\geq$14~kpc), as discussed in Sect.~\ref{NGC4579-potential}.

In the following, we will suppose that the corotation resonance lies at the end of the bar: $r_{CR}\sim$6$\pm$1~kpc, i.e.,  allowing for a 20$\%$ uncertainty in $r_{CR}$ . This is a plausible scenario for the well-developed although relatively weak bar of NGC\,4579  (see Sect.~\ref{grav}). An estimate of the bar pattern speed ($\Omega_{BAR}$) is thus obtained using the $\Omega$ curve derived from CO if we rely on the epicyclic approximation; this yields $\Omega_{BAR}$\,$\sim$50$\pm$10~km~s$^{-1}$kpc$^{-1}$ (See Fig.~\ref{fig:frequencies}). The two ILRs lie approximately at $r$\,$\sim$500~pc (inner ILR=iILR) and $r$\,$\sim$1.3~kpc (outer ILR=oILR) for $\Omega_{BAR}$\,$\sim$50~km~s$^{-1}$kpc$^{-1}$ (Fig.~\ref{fig:frequencies}). This value of $\Omega_{BAR}$ allows the bar to have an extended ILR region, a requirement imposed by the observed distribution and kinematics of molecular gas in the central $r$\,$\sim$2~kpc disk. In the context of this approximation we can put an upper limit to the pattern speed of the bar of NGC\,4579: $\Omega_{BAR}$\,$\sim$60~km~s$^{-1}$kpc$^{-1}$; this value of $\Omega_{BAR}$ would impose the existence of a single ILR at $\sim$800~pc, a scenario incompatible with observations, which require the existence of two ILRs. With $\Omega_{BAR}$\,$\sim$50~km~s$^{-1}$kpc$^{-1}$, the O-ring would correspond to the oILR, a resonance which is currently being depopulated due to gas inflow down to a region inside the iILR of the bar (see discussion in Sect.~\ref{grav-results}). The latter explains the molecular gas concentration in the inner spiral arms (N-arm and S-arm), formed as a mixture of x$_2$ and x$_1$ precessing orbits. On its way to the nucleus, molecular gas is forming stars in the inner spiral arms, as evidenced by the detected H$_{\alpha}$ emission, more intense in the case of the N-arm (see Fig.~\ref{fig:overlay Ha-CO}), as discussed in Sect.~\ref{tracers-SF}.

\subsection{The nuclear oval}\label{tracers-oval}

The $K$-band image of NGC\,4579 shows evidence of a previously unnoticed oval distortion present
in the inner $r$\,$\sim$200~pc (Fig.~\ref{fig:overlay UV-CO21}). This nested structure was not detected in the $I$-band HST image originally used in the first gravity torque analysis of NGC\,4579 (GB05). The oval is oriented along $PA'_{oval}$\,$\sim$70$^{\circ}$ (measured from +X axis in the plane of the galaxy), and therefore is not aligned with the large-scale bar. Misalignment can be taken as evidence of decoupling of the nested structure. The reality of this distortion is confirmed by the analysis of the Fourier decomposition of the potential described in Sect.~\ref{grav}. Fig.~\ref{fig:overlay UV-CO21} shows a zoomed view of the central $r$\,$\sim$200~pc-region. The oval appears encircled by an off-centered ellipsoidal ring  detected in the HST UV image of the galaxy, first discussed by Contini~(\cite{con04}). The UV ring has an average radius $r$\,$\sim$150-200~pc and has been recently described by Comer\'on et al.~(\cite{come08}) as an ultra-compact nuclear ring. This ring can be interpreted as a resonance of the oval distortion. The best guess assigns the Ultra Harmonic Resonance (UHR) of the oval to the UV ring: if the pattern speed of the oval is $\Omega_{OVAL}\sim$270$\pm$10~km~s$^{-1}$kpc$^{-1}$, the UHR of the oval should lie at $\sim$200~pc and its corotation
at $\sim$1~kpc (Fig.~\ref{fig:frequencies}). With this choice we can explain the decoupling of the oval with respect to the large-scale bar, as the corotation of the oval would be close to the oILR of the large-scale bar. This resonance overlap is a prerequisite to the decoupling of nuclear bars/ovals (e.g., Garc\'ia-Burillo et al.~\cite{gb98}; Hunt et al.~\cite{hun08}).

As discussed in Sect.~\ref{grav-results}, the oval perturbation is key to favoring gas fueling to the AGN at present.
More particularly, the decoupling scenario explains the depopulation of the oILR of the large-scale bar, under the 
combined action of the bar and the nuclear oval. A similar time sequence is nicely illustrated by the numerical simulations adapted to follow the nuclear bar decoupling in the double-barred galaxy NGC\,2782, discussed by Hunt et al.~(\cite{hun08}).  

%%%%%%%%%%%%%%%%%%%%%%%%%%%%%%%%%%%%%%%%%%%%%%%%%%%%%%%%%%%%%%%%%%%%%%%%%%%%%%%%%%%%%%%%%%%%%%%%

\begin{figure*}[tbh!]
   \centering
   \includegraphics[width=13.5cm]{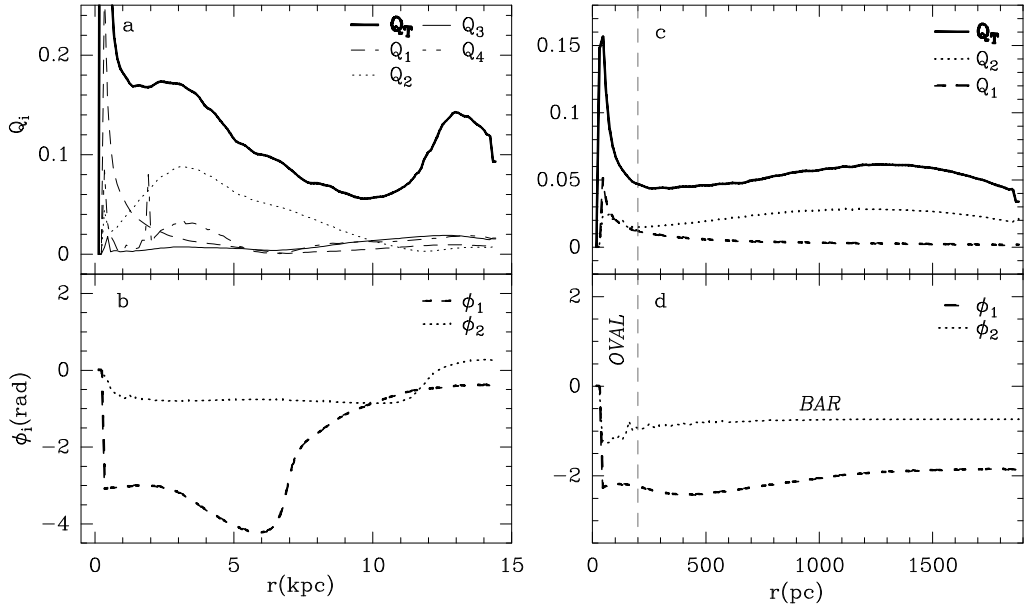}
   \caption{Strength ($Q_{1}$, $Q_{2}$ and $Q_{T}$) and phases ($\phi_{1}$ and $\phi_{2}$) of $m=1$ and $m=2$ Fourier components of the stellar potential derived from the $K$-band WHT image inside $r$\,$\sim$15~kpc ({\it left panels}: a, b) and inside $r$\,$\sim$2~kpc ({\it right panels}: c, d). We highlight the radii where the OVAL and the BAR perturbations prevail at $r$\,$\leq$2~kpc. In addition, $Q_{3}$, and $Q_{4}$ values are plotted in panel a to illustrate the predominance of the bar mode in the disk up to $r$\,$\sim$7--8~kpc.}
   \label{fig:potential}
 \end{figure*}

%%%%%%%%%%%%%%%%%%%%%%%%%%%%%%%%%%%%%%%%%%%%%%%%%%%%%%%%%%%%%%%%%%%%%%%%%%%%%%%%%%%%%%%%%%%%%%%%%%

\subsection{The gas distribution in the outer disk ($r>$4~kpc)}\label{tracers-outer}

The 21cm HI map of NGC\,4579 gives a sharp view of the gas distribution in the outer disk of the galaxy (from $r$\,$\sim$4 to 15~kpc) (See Figs.~\ref{fig:pot_deproj}\textit{a} and ~\ref{fig:overlay Spitzer/IRAC-HI}). The 8$\mu$m Spitzer/IRAC and the FUV GALEX images provide a detailed picture of the star formation in the disk. For a Sb galaxy like NGC\,4579, $>$70$\%$ of the total emission at 8$\mu$m is of non-stellar origin and is produced by warm dust heated by star formation activity (e.g., Pahre et al.~\cite{pah04}).  The HI distribution in the disk shows a central depression in the inner $r$\,$\sim$4~kpc, mimicking a ring. The HI pseudo-ring is partly incomplete and has an average diameter of $\sim$12~kpc. At close sight, the HI pseudo-ring appears to be the superposition of two winding spiral arms. The inner segments of the spiral arms are better identified in the 8$\mu$m Spitzer/IRAC and the GALEX images; the 8$\mu$m and FUV arms neatly delineate the inner edge of the HI pseudo-ring. Further out, the arms unfold in the disk over 180$^{\circ}$ in azimuth and end up forming the pseudo-ring detected in HI. Most remarkably, the spiral arms are not connected to the two bar ends located at ($r$,$\Phi$)=(6\,~kpc,\,~45$^{\circ}$) and (6\,~kpc,\,~225$^{\circ}$). Instead, the two-arm spiral structure starts at a different azimuth ($\sim$90$^{\circ}$ from the bar ends) and, also, at a significantly smaller radius in the disk: ($r$,$\Phi$)=(4\,~kpc, 135$^{\circ}$) and (4\,~kpc, 315$^{\circ}$). 

On the whole, the particular geometry described above suggests that the bar and the spiral are decoupled and thus do not share a common pattern speed. It is accepted that spirals and bars can rotate at differing angular speeds in barred spiral galaxies, a scenario first advocated by Sellwood \& Sparke~(\cite{sel88}) who contemplate that, under decoupling, outer spirals can rotate at lower $\Omega_p$. As recently shown by numerical simulations (e.g., Combes~\cite{com07} and references therein), decoupling is more likely in the case of weak bars, of which the NGC\,4579's bar is a good example. As discussed in Sect.~\ref{grav-results}, this situation favors gas fueling to the central regions, as the corotation barrier imposed by the bar could be thus overcome. The morphology of the outer disk revealed by HI in NGC\,4579 suggests that the neutral gas is currently piling up in an inner ring located well inside the corotation of the bar. This scenario is confirmed by the coincidence of the inner edge of the HI pseudo-ring with the predicted position of the UHR of the bar, around $r_{UHR}$\,$\sim$3.8~kpc, as illustrated in Fig.~\ref{fig:frequencies}. 

\subsection{The star formation rate in NGC\,4579}\label{tracers-SF}

The star formation pattern in the inner disk ($r\leq$2~kpc) of NGC\,4579 have been previously described 
by Pogge et al.~(\cite{pog00}) and Contini~(\cite{con04}), based on HST optical and UV images (see also
descriptions by GB05 and Comer\'on et al.~\cite{come08}). Most of the H$_{\alpha}$ emission in the region 200~pc$<r<$2~kpc comes from a two-arm spiral structure that shows a geometry similar to that of the CO {\it inner spiral arms}, described in Sect.\ref{CO-distribution} (see Fig.~\ref{fig:overlay Ha-CO}). In addition, part of the H$_{\alpha}$ emission is related to the CO {\it outer ring}. Closer to the AGN ($r<$200~pc), the bulk of the H$_{\alpha}$ emission is seen to come from a slightly off-centered disk. At close sight, several subcomponents can be identified in this disk. There is a point-like nuclear source of $\sim$0.27$\arcsec$-size, located at the dynamical center of the galaxy; the emission of this component has been interpreted as due to photo-ionization by the AGN (Maoz et al.~\cite{mao05}).  Furthermore, the OIII line and UV maps reveal also a one-sided fan-shaped morphology in the emission in the inner $r$\,$\sim$100~pc of NGC\,4579 (Pogge et al.~\cite{pog00}; Contini~\cite{con04}). On larger scales, the HST UV image of the galaxy, shows an off-centered ring at $r$\,$\sim$150-200~pc (see Sect.~\ref{tracers-oval} and Comer\'on et al.~\cite{come08}). As described in Sect.~\ref{tracers-outer}, star formation in the outer disk, as traced by the UV continuum emission seen by GALEX, arises from a pseudo-ring characterized by a geometry similar to that of the HI pseudo-ring.

In order to quantify the amount of gas transformed into stars, we have used
the HST H$_{\alpha}$ and GALEX UV images described above to calculate the corresponding star formation rates (SFR) on different regions throughout the disk of the galaxy. SFR values are derived using the empirical formulae given by Kennicutt~(\cite{ken98}). We first estimate the SFR in the inner disk; more precisely, we define SFR$_{inner}$ as the integrated SFR inside the region 300~pc$<$\,$r$\,$<$2~kpc. To correct for extinction, and thus have a more accurate evaluation of SFR$_{inner}$, we have used the CO(1--0) PdBI map to derive the H$_2$ column density map, using the CO-to-H$_2$ conversion factor of Solomon \& Barrett~(\cite{sol91}): $X$=$N(H_2)/I_{CO}$=2.3$\times$10$^{20}$~cm$^{-2}$~K$^{-1}$~km$^{-1}$~s. The A$_V$ map is then derived using the commonly-used N(H$_2$)--to--A$_v$ conversion factor of Bohlin et al~(\cite{boh78}): N(H$_2$)/A$_V$=10$^{21}$cm$^{-2}$mag$^{-1}$. Adopting a screen-geometry for the obscuring dust in front of the HII regions, we must scale the A$_v$ map by a factor 1/2, if HII regions are assumed to lie in the mid-plane of the galaxy. This scaled extinction map, with A$_v$ typically ranging from $\sim$3 to $\sim$15, is used to derive A$_{H_{\alpha}}$, following Cardelli et al.~(\cite{car89}): A$_{H_{\alpha}}$=0.828$\times$A$_v$. We derive SFR$_{inner}$\,$\sim$0.10~M$_{\sun}$~yr$^{-1}$, confirming that the nuclear starburst of NGC~4579 is only moderate. In a second step, we use SFR$_{inner}$ to calculate the SFR throughout the disk of the galaxy by scaling the integrated fluxes measured in the GALEX UV image at the corresponding radii. In particular, we define SFR$_{outer}$ as the SFR integrated inside the region 4~kpc$<$\,$r$\,$<$10~kpc. We derive SFR$_{outer}$\,$\sim$1.30~M$_{\sun}$~yr$^{-1}$. 

As such, SFR$_{inner}$ and  SFR$_{outer}$ contain the bulk of the total SFR of NGC\,4579 estimated as
SFR$_{total}$\,$\sim$1.70~M$_{\sun}$~yr$^{-1}$. These SFR are compared to the corresponding inflow rates driven by gravity torques in these regions, relevant to understand the overall gravity torque budget, as discussed in Sect.~\ref{NGC4579-mass}.

\section{Gravity torques}\label{grav}

\subsection{Methodology}\label{method}

Firstly we have derived a good representation of the stellar potential for the disk of NGC\,4579. To do so we have used the $K$-band image of the galaxy published by Knapen et al.~(\cite{kna03}), assuming that this image is virtually free from dust extinction or stellar population biases. This working hypothesis better applies to the $K$-band image than to the $I$-band image originally used by GB05 to derive the gravity torque budget in NGC\,4579. The $K$-band image has been deprojected according to the angles $PA$=95~$^{\circ}$ and $i$=36~$^{\circ}$, assumed in Sect.~\ref{introduction}. Figure.~\ref{fig:pot_deproj} shows the overlay of the deprojected gas distribution as derived from the CO and HI maps with the deprojected $K$-band image of NGC\,4579.

To reproduce the vertical mass distribution, the $K$-band image was convolved using an isothermal plane model of constant scale height, equal to $\sim$1/12th of the radial scale-length of the image. We also adopted a constant mass-to-light (M/L) ratio, obtained by fitting the rotation curve to the CO and HI data. The potential--$\Phi(R,\theta)$--was then calculated by a Fourier transform method and decomposed in the different m-modes as follows:

\begin{equation}
\Phi(R,\theta) = \Phi_0(R) + \sum_m \Phi_m(R) \cos (m \theta - \phi_m(R))
\end{equation}

\noindent
where $\Phi_m(R)$ and $\phi_m(R)$ are the amplitude and phase of the m-mode, respectively.

According to Combes \& Sanders~(\cite{com81}), the strength of the $m$-Fourier component is defined as:
$Q_m(R)$ as
 \begin{equation}
Q_m(R)=m \Phi_m / R | F_0(R) |
 \end{equation}

Similarly, the strength of the total non-axisymmetric perturbation is obtained by:
\begin{equation}
Q_T(R) = {F_T^{max}(R) \over F_0(R)} =
{{{1\over R}\bigl{(}{\partial \Phi(R,\theta)\over \partial\theta}\bigr{)}_
{max}} \over {d\Phi_0(R)\over dR}}
\end{equation}
\noindent
where $F_T^{max}(R)$ represents the maximum amplitude of the tangential force over all $\theta$ and $F_{0}(R)$ is the mean axisymmetric radial force.

Figure~\ref{fig:potential} shows the description of 
the gravitational potential of NGC~4579 defined by the Q$_{i=1,2}$, Q$_T$, and $\phi_{i=1,2}$ curves, used in the discussion of Sect.~\ref{NGC4579-potential}. 

%%%%%%%%%%%%%%%%%%%%%%%%%%%%%%%%%%%%%%%%%%%%%%%%%%%%%%%%%%%%%%%%%%%%%%%%%%%%%%%%%%%%%%%%%%%%%%%%

\begin{figure}[tbh!]
   \centering
    \includegraphics[width=8.5cm]{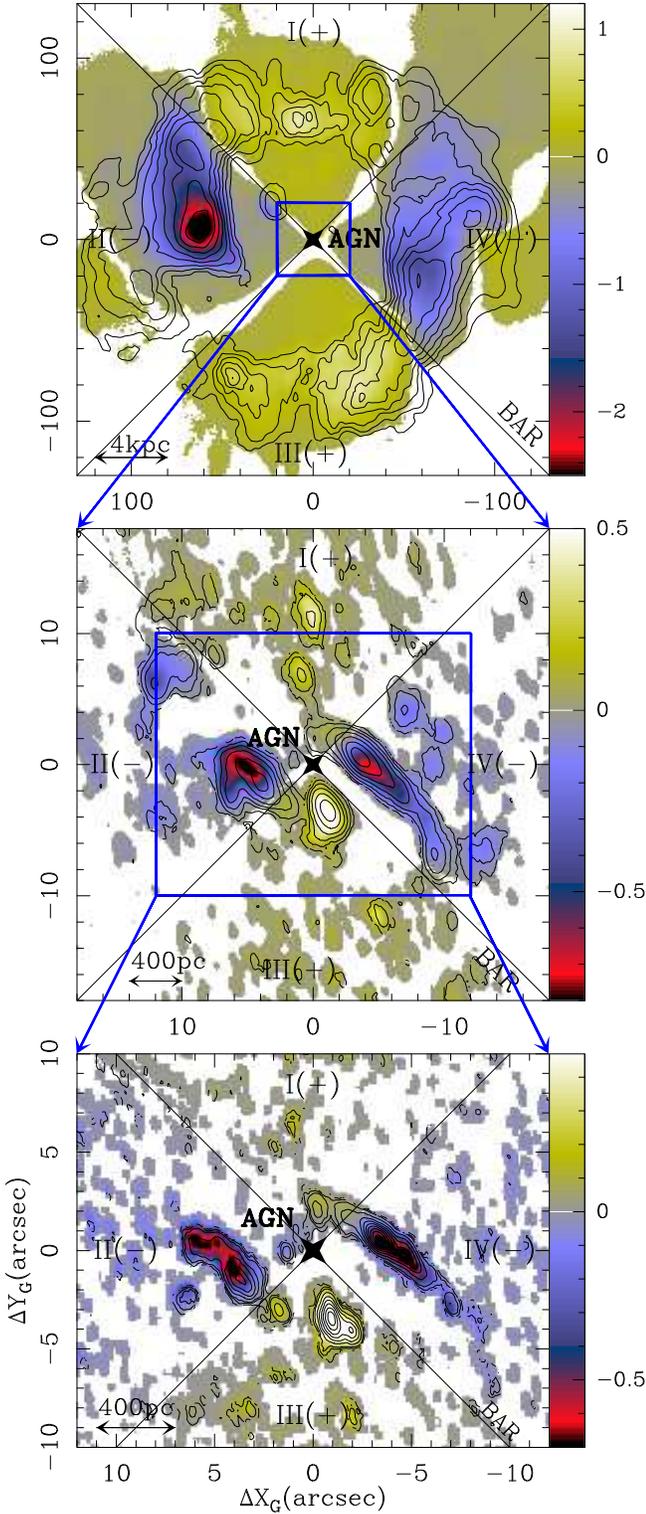}
   \caption{{\bf a)}({\it Upper panel})~We overlay the HI contours with the the map of the (dimensionless) effective angular momentum variation, as defined in text. The derived torques change sign as expected, if the {\it butterfly} diagram, defined by the orientation of quadrants I-to-IV, can be attributed to the action of the large-scale bar. {\bf b)}({\it Middle panel})~ The same as {\bf a)} but for CO(1--0). {\bf c)}({\it Lower panel})~The same as {\bf a)} but for CO(2--1). Maps are deprojected onto the galaxy plane.}
   \label{fig:torques-HI-CO}
 \end{figure}

%%%%%%%%%%%%%%%%%%%%%%%%%%%%%%%%%%%%%%%%%%%%%%%%%%%%%%%%%%%%%%%%%%%%%%%%%%%%%%%%%%%%%%%%%%%%%%%%%%%

The forces per unit mass ($F_x$ and $F_y$) are calculated from the derivatives of $\Phi(R,\theta)$ at 
each pixel; the gravity torques per unit mass--$t(x,y)$--can be then computed by:

\begin{equation}
t(x,y) = x~F_y -y~F_x
\end{equation}

The sense of the circulation of the gas in the galaxy plane
(counterclockwise in NGC\,4579; see Sect.~\ref{CO-kinematics}) defines whether the sign of $t(x,y)$ is positive (negative) if the torque accelerates (decelerates) locally  the gas in the disk.
Note that the specific torque field $t(x,y)$ does not depend on the present distribution of the gas. However, we use the $t(x,y)$ map to infer the angular momentum variations by assuming that the gas column density ($N(x,y)$) derived from the CO line maps (in the inner disk) and from the HI map (in the outer disk) is a good estimate of the probability of finding gas particles at this location at present. The torque field is then weighted by $N(x,y)$ at each pixel to derive the time derivative of the angular momentum surface density in the galaxy plane, $dL_s(x,y)/dt$=$N(x,y) \times t(x,y)$. Our approach is statistical since we do not follow particles along individual orbits, at best estimated through model fitting (e.g., see Boone et al.~\cite{boo07}), in order to derive angular momentum variations. Instead, we average over all orbits of gas particles at each position and account for the time spent by the gas clouds along all the possible orbit paths. Figure~\ref{fig:torques-HI-CO} shows the dimensionless version of the torque maps normalized by [$\vert$~$N(x,y)$~$\times$~$t(x,y)$~$\vert$]$_{max}$, as derived from CO and HI. Sect.~\ref{NGC4579-transfer} provides a detailed description of these maps.

To estimate the net radial gas flow globally induced by the torque field in the galaxy disk, we have computed the torque per unit mass averaged over the azimuth $\theta$ as a function of radius $R$, using $N(x,y)$ as weighting function:

\begin{equation}
t(R) = \frac{\int_\theta N(x,y)\times(x~F_y -y~F_x)}{\int_\theta N(x,y)}
\end{equation}

By definition, $t(R)$ is the azimuthally-averaged time derivative of the specific angular momentum--$L$--of the gas, i.e., $t(R)$=$dL/dt$\,$\vert_\theta$.  Similarly to the
torque maps, the sign of $t(R)$, either positive or negative, defines whether the gas gains or loses angular momentum, respectively.  Specifically, we estimate the AGN feeding efficiency by deriving the average fraction of the gas specific angular momentum transferred in one rotation (T$_{rot}$) by the stellar potential, as a function of radius, i.e., by the dimensionless function $\Delta L/L$ defined as:

\begin{equation}
{\Delta L\over L}=\left.{dL\over dt}~\right\vert_\theta\times \left.{1\over L}~\right\vert_\theta\times 
T_{rot}={t(R)\over L_\theta}\times T_{rot}
\label{tgrav}
\end{equation}

\noindent
where $L_\theta$ is assumed to be well represented by its axisymmetric average, $L_\theta=R\times v_{rot}$. 
The absolute value of $L/\Delta L$ determines how much time (in T$_{rot}$ units) the stellar potential will need to transfer the entirety of the gas angular momentum. A small
value of $\Delta L / L$ implies that the stellar potential is inefficient at present. The $\Delta L/L$ curves derived from the CO and HI maps of NGC~4579, shown in Fig.~\ref{fig:torques-radial-HI-CO}, are used in the discussion of Sect.~\ref{NGC4579-transfer}.

Following the approach of GB05, we can estimate the gas mass inflow (- sign)/outflow (+ sign) rate driven by the stellar potential per unit length as a function of radius (in units of M$_{\odot}$~yr$^{-1}$pc$^{-1}$ in 
Figs.~\ref{fig:mass-radial-HI}\textit{a} to \ref{fig:mass-radial-CO21}\textit{a}) according to: 

\begin{equation}
{d^2M\over dRdt}=\left.{dL\over dt}~\right\vert_\theta\times \left.{1\over L}~\right\vert_\theta\times 2\pi 
R\times \left.N(x,y)~\right\vert_\theta
\end{equation}
\noindent
where $\left.N(x,y)~\right\vert_\theta$ is the radial profile of $N(x,y)$ averaged over the azimuth for a radial binning $\Delta R$, derived from HI, for the outer disk, and from CO, for the inner disk. 
 
Similarly, the inflow/outflow rates integrated out to a certain radius R can be derived as:
\begin{equation}
{dM\over dt}=\sum {d^2M\over dRdt}~\times\Delta R
\end{equation}

Figs.~\ref{fig:mass-radial-HI}\textit{b} to \ref{fig:mass-radial-CO21}\textit{b} display these integrated rates in units of M$_{\odot}$~yr$^{-1}$. The mass transfer rates in NGC~4579 are discussed in Sect.~\ref{NGC4579-mass}.

\subsection{The gravity torque budget in NGC\,4579}\label{grav-results}

\subsubsection{The stellar potential}\label{NGC4579-potential}

To derive the stellar potential $\Phi(R,\theta)$ used in the gravity torque calculation of NGC\,4579, we have followed an adaptive approach.  $\Phi(R,\theta)$ in the inner $r$\,$\sim$\,2~kpc disk has been derived using a grid of 256x256 pixels with 0.15$\arcsec$~pixel$^{-1}$ on the $K$-band image. This representation of $\Phi(R,\theta)$ is the best adapted to derive the torque field on CO. On the other hand, a spatially smoothed version of the NIR image, with 256x256 pixels and 1.15$\arcsec$~pixel$^{-1}$, has been used to calculate $\Phi(R,\theta)$ in the outer disk up to $r$\,$\sim$15~kpc. The ensuing torque field on HI is derived from this representation of the stellar potential.

Figure~\ref{fig:potential} shows the Q values derived as a function of radius. The large-scale bar ($m=2$) is clearly detected in the smoothed version of $\Phi(R,\theta)$ (Figs.~\ref{fig:potential}\textit{a,b}). The area of influence of the bar, determined by the dominance of the $m=2$ mode over other secondary modes extends up to $r$\,$\sim$7--8~kpc in the disk. For $r$\,$\geq$8~kpc, $Q_1$+$Q_3$+$Q_4\geq$\,$Q_2$, an indication that the bar mode is not dominant on these scales of the outer disk. This puts an upper limit to the size of the bar ($r$\,$<$7--8kpc). As discussed in Sect.~\ref{tracers-bar}, the isophotal analysis of the NIR images of the galaxy gives a smaller size for the bar: $r$\,$\sim$5--6~kpc. The finally adopted value for $r_{CR}$=6$\pm$1~kpc is a reasonable compromise between the two determinations above and it accounts for the overall gas response in the disk.

On the whole, the bar is moderately strong: the $Q_2$ strength of the bar is maximum at $r$\,$\sim$3~kpc where it reaches a value of $\sim$0.1. Moreover, the phase of the bar $\phi_2$ stays noticeably constant at a value $\sim$--0.8 radians, i.e., at an angle $\sim$45$^{\circ}$, measured counterclockwise from +X-axis in Figs.~\ref{fig:pot_deproj}. This also confirms the value of $PA'_{bar}$ derived from the isophotal analysis of Sect.~\ref{tracers-bar}. The inner representation of $\Phi(R,\theta)$ also reveals the bar at $r$\,$<$2~kpc (Figs.~\ref{fig:potential}\textit{c,d}), with a similarly constant phase and, not surprisingly, with an average lower strength, $Q_2$\,$\sim$0.02--0.03 (at these radii we are inside the peak of $Q_2$). Of particular note, the high-resolution version of $\Phi(R,\theta)$ unveils the oval distortion with a size and an orientation similar to those derived in Sect.~\ref{tracers-oval}. The presence of a misaligned oval is betrayed by a significant $\sim$25$^{\circ}$ change in the phase of the $m$=2 mode. $\phi_2$ gradually shifts from the large-scale bar orientation, $\sim$--0.8 radians for 200~pc$<$\,$r$\,$<$7~kpc ($\sim$45$^{\circ}$), to $\sim$--1.2 radians ($\sim$70$^{\circ}$) at $r$\,$\sim$100~pc.

%%%%%%%%%%%%%%%%%%%%%%%%%%%%%%%%%%%%%%%%%%%%%%%%%%%%%%%%%%%%%%%%%%%%%%%%%%%%%%%%%%%%%%%%%%%%%%%%%%%

\begin{figure}[tbh!]
   \centering
   \includegraphics[width=8.5cm]{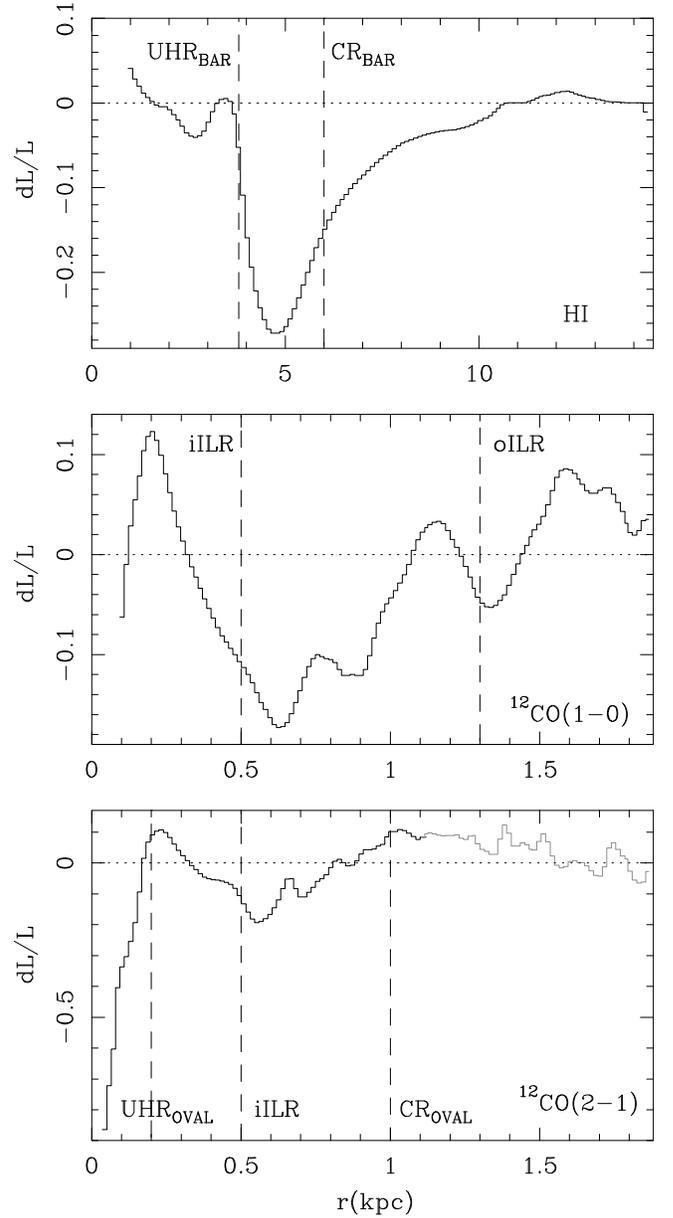}
   \caption{The average fraction of the angular momentum transferred from/to the gas in one rotation--$dL/L$--are plotted as a function of radius, as derived from the HI ({\it upper panel}), CO(1--0) ({\it middle panel}) and CO(2--1) ({\it lower panel}) maps of the disk of NGC\,4579. The locations of BAR resonances (iILR, oILR, CR$_{BAR}$, UHR$_{BAR}$) and those of OVAL resonances (CR$_{OVAL}$, UHR$_{OVAL}$) are highlighted. We have blocked the central $r$\,$\sim$1~kpc in the $dL/L$ plot obtained from the HI map, due to the spatial resolution limit of the HI data. For similar reasons, the central $r$\,$\sim$100~pc and $r$\,$\sim$50~pc around the AGN are blocked in the $dL/L$ plots obtained from the CO 1--0 and 2--1 maps, respectively. In the range 1.8~kpc$>$\,$r$\,$>$1.2~kpc, estimates of $dL/L$ based on the CO(1--0) map are more reliable than those derived from CO(2--1) data (in grey), as the latter come from a region outside the PdBI primary beam at this frequency. The same display conventions are applied to Figs.~\ref{fig:mass-radial-HI} to \ref{fig:mass-radial-CO21}.}
\label{fig:torques-radial-HI-CO}
 \end{figure}

%%%%%%%%%%%%%%%%%%%%%%%%%%%%%%%%%%%%%%%%%%%%%%%%%%%%%%%%%%%%%%%%%%%%%%%%%%%%%%%%%%%%%%%%%%%%%%%%%%%
%%%%%%%%%%%%%%%%%%%%%%%%%%%%%%%%%%%%%%%%%%%%%%%%%%%%%%%%%%%%%%%%%%%%%%%%%%%%%%%%%%%%%%%%%%%%%%%%%%%

\begin{figure}[tbh!]
   \centering
   \includegraphics[width=8.5cm]{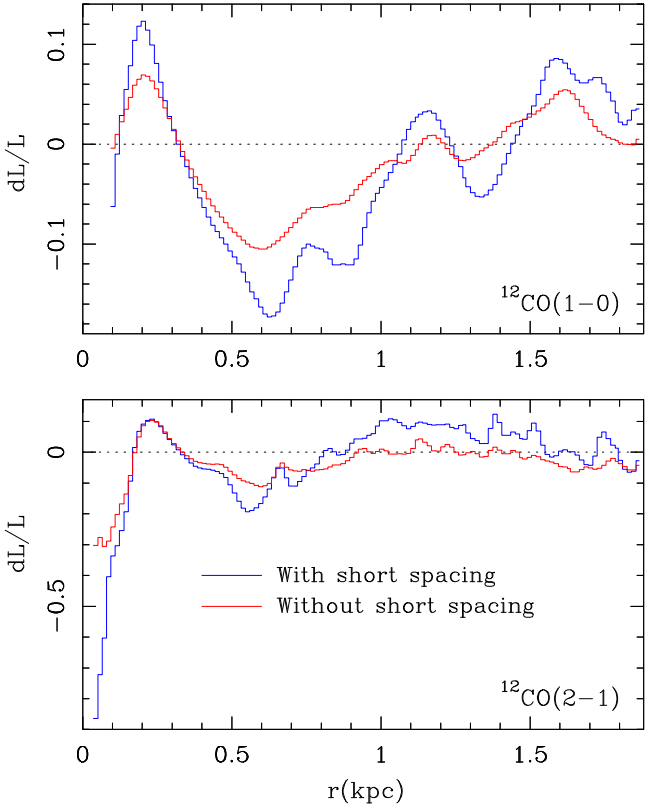}
   \caption{Same as Fig.~\ref{fig:torques-radial-HI-CO}, but here showing the results obtained for $dL/L$ from CO
maps with and without short spacing correction.}
\label{fig:torques-radial-comparison}
 \end{figure}

%%%%%%%%%%%%%%%%%%%%%%%%%%%%%%%%%%%%%%%%%%%%%%%%%%%%%%%%%%%%%%%%%%%%%%%%%%%%%%%%%%%%%%%%%%%%%%%%%%%

There are discrepancies between the stellar potential obtained from the $K$-band image of NGC~4579, and the  representation of $\Phi(R,\theta)$ obtained by GB05 from the $I$-band HST image of the galaxy. Firstly, the strong $m=1$ mode detected in the $I$-band image is much weaker in the new evaluation of $\Phi(R,\theta)$. As illustrated in Fig.~7 of GB05, the $m=1$ mode seemed to overtake the potential strength of the bar at $r$\,$<$500~pc: $Q_1$ was seen to vary between $\sim$0.02 and $\sim$0.10, while 
$Q_2$\,$\sim$0.02. This is in contrast to the new result shown in Fig.~\ref{fig:potential}\textit{c}: $Q_2>Q_1$, except for the inner $r$\,$\sim$50~pc, i.e., below the limit of the spatial resolution of the image\footnote{The peak of $Q_1$ at r$\sim$50--100~pc identified in Fig.~\ref{fig:potential}\textit{a} is an artifact created by the spatial smoothing performed to derive the large-scale representation of $\Phi(R,\theta)$, and as such is unreliable.}. 
This has consequences on the interpretation of the lopsidedness identified in the gas
distribution and kinematics probed by CO. Lopsidedness cannot be clearly attributed to the action of a {\it true} $m=1$ mode in the stellar potential. As a second important difference, the oval identified in the $K$-band image was not present in the $I$-band image of GB05. The reported discrepancies in $\Phi(R,\theta)$ can be attributed to the residual effects of extinction which are noticeable in the central $r$\,$\sim$2~kpc of the $I$-band image.

Taken together, these differences are relevant to explaining the new version of the gravity torque budget in NGC\,4579, described in Sect.~\ref{NGC4579-transfer}.

\subsubsection{Gas-flow time-scales in NGC~4579}\label{NGC4579-transfer}

Figure~\ref{fig:torques-HI-CO} shows the 2D pattern of gravitational torques obtained throughout the disk of NGC\,4579, from $r$\,$\sim$\,15~kpc down to the inner $r$\,$\sim$50\,pc around the AGN. The derived torques change sign following a characteristic 2D {\it butterfly} pattern. Quadrants I-to-IV will define hereafter the regions where the signs of the torques driven by the dominant perturbation of the stellar potential are expected to be constant.

Assuming that the gas is rotating counterclockwise, 
Fig.~\ref{fig:torques-HI-CO}\textit{a} shows that, although HI emission is distributed along the four 
quadrants, the majority of the gas is seen to lie along the leading edges of the large-scale bar where torques are negative (quadrants II(-) and IV(-)), especially on intermediate scales (4~kpc$<$\,$r$\,$<$8~kpc). Figure~\ref{fig:torques-radial-HI-CO}\textit{a} shows that the azimuthally averaged torques are systematically negative from $r$\,$\sim$10~kpc down to $r$\,$\sim$4~kpc. This result suggests that the corotation barrier of the bar in NGC\,4579 (at $\sim$6~kpc) has been overcome due to secular evolution processes. If the gas response was the canonical spiral pattern coupled to the bar, the corotation would separate in the disk the regions where torques are positive (from corotation to the Outer Lindblad Resonance (OLR)) from those where torques are negative (from corotation to the oILR). Positive torques would tend to form a ring at the OLR. However, a decoupling of the spiral allows the gas to efficiently populate the UHR region inside corotation and thus produce net gas inflow on intermediate scales in the disk. A similar scenario has been suggested to explain the gravity torque maps of some of the NUGA galaxies recently analyzed by Haan et al.~(\cite{haa08b}). The implied time-scales for angular momentum transfer in NGC~4579 are $\sim$5--10 rotation periods. On its way to the nucleus, the gas stops in the outer disk close to the UHR of the bar, where torques become negligible (Fig~\ref{fig:torques-radial-HI-CO}\textit{a}).

In the inner disk, Fig.~\ref{fig:torques-HI-CO}\textit{b,~c} shows that the bulk of the CO emission
along the inner spiral arms lies where torques are strong and negative (quadrants II(-) and IV(-)). The corresponding azimuthally averaged torques, shown in Fig~\ref{fig:torques-radial-HI-CO}\textit{b,~c}, are mostly negative from $r$\,$\sim$1.3~kpc down to $r$\,$\sim$300~pc. This indicates a net gas inflow from the oILR down to a region inside the iILR of the bar. As for the outer disk, the implied time-scales for angular momentum transfer are also $\sim$5--10 rotation periods. This result is in agreement with our first calculation published by GB05 (see Fig.~11 of GB05) regarding the sign of the torques. However the gravity torques shown in Fig~\ref{fig:torques-radial-HI-CO}\textit{b,~c} are comparatively stronger at the same radii, in particular those contributing to gas inflow.  The changing picture can be attributed to the new $\Phi(R,\theta)$ used in this work but, also, to the new $N(x,y)$--values that now include the short spacing correction for CO. This is illustrated in 
Fig.~\ref{fig:torques-radial-comparison} where we compare the $dL/L$ curves obtained with the new $\Phi(R,\theta)$ used in this work, but with two different versions of $N(x,y)$, derived from the PdBI-only CO data (used by GB05), and from the short spacing corrected PdBI+30m data (this work), respectively. In summary, we find that the gas in the inner disk is funneled from the oILR to a region inside the iILR at $r$\,$\sim$300~pc. There is a barrier of positive torques between $r$\,$\sim$200\,pc (i.e., close to the UHR of the oval perturbation) and $r$\,$\sim$300\,pc (Fig~\ref{fig:torques-radial-HI-CO}\textit{b,~c}).

Closer to the AGN ($r$\,$<$200~pc), the bulk of the CO emission is concentrated north of the nucleus (complex N) in the trailing quadrant of the large-scale bar I(+), where stellar torques are positive. However, the rest of the 
central disk complexes from $r$\,$\sim$50\,pc to $r$\,$\sim$150\,pc, identified in the CO(2--1) map, feel negative torques due to the combined action of the large-scale bar and the inner oval. The two $m=2$ modes act in concert to produce net gas inflow down to the spatial resolution of our observations (see Fig~\ref{fig:torques-radial-HI-CO}\textit{c}).
This result is different to that reported in GB05, where the evidence of AGN fueling at $r$\,$<$200~pc was lacking.  
The implied time-scales for angular momentum transfer are extremely short: $\sim$1--3 rotation periods.

%%%%%%%%%%%%%%%%%%%%%%%%%%%%%%%%%%%%%%%%%%%%%%%%%%%%%%%%%%%%%%%%%%%%%%%%%%%%%%%%%%%%%%%%%%%%%%%%%%%

\begin{figure}[tbh!]
   \centering
   \includegraphics[width=8.5cm]{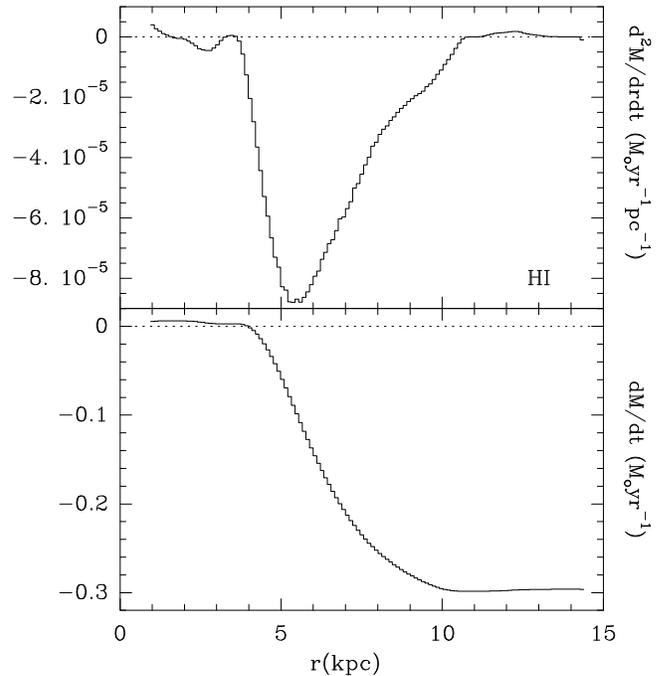}
   \caption{{\bf a)}({\it Upper panel}) We represent the radial variation of the mass inflow(-) or outflow(+) rate of gas per unit radial length ($d^2M/dRdt$) in the nucleus of NGC\,4579 due to the action of stellar gravitational torques on the gas distribution as derived from HI data. Units are M$_{\sun}$~yr$^{-1}$pc$^{-1}$.
   {\bf b)}({\it Lower panel}) We plot the mass inflow/outflow rate integrated inside a certain radius $r$ ($dM/dt$) in 
M$_{\sun}$~yr$^{-1}$. }
\label{fig:mass-radial-HI}
 \end{figure}

%%%%%%%%%%%%%%%%%%%%%%%%%%%%%%%%%%%%%%%%%%%%%%%%%%%%%%%%%%%%%%%%%%%%%%%%%%%%%%%%%%%%%%%%%%%%%%%%%%%

%%%%%%%%%%%%%%%%%%%%%%%%%%%%%%%%%%%%%%%%%%%%%%%%%%%%%%%%%%%%%%%%%%%%%%%%%%%%%%%%%%%%%%%%%%%%%%%%%%%

\begin{figure}[tbh!]
   \centering
   \includegraphics[width=8.5cm]{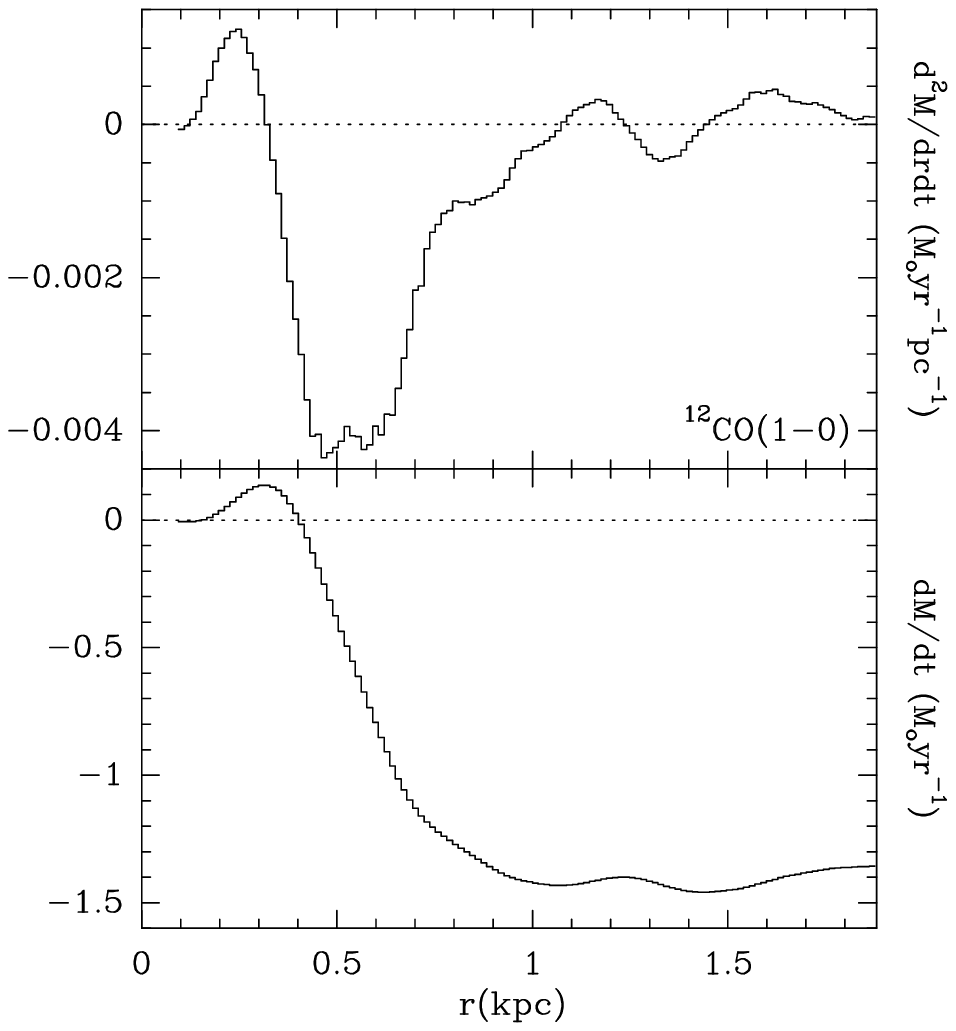}
   \caption{Same as Fig.\ref{fig:mass-radial-HI} but here based on the gas distribution as derived from CO(1--0) data.}
\label{fig:mass-radial-CO10}
 \end{figure}

%%%%%%%%%%%%%%%%%%%%%%%%%%%%%%%%%%%%%%%%%%%%%%%%%%%%%%%%%%%%%%%%%%%%%%%%%%%%%%%%%%%%%%%%%%%%%%%%%%%

%%%%%%%%%%%%%%%%%%%%%%%%%%%%%%%%%%%%%%%%%%%%%%%%%%%%%%%%%%%%%%%%%%%%%%%%%%%%%%%%%%%%%%%%%%%%%%%%%%%

\begin{figure}[tbh!]
   \centering
   \includegraphics[width=8.5cm]{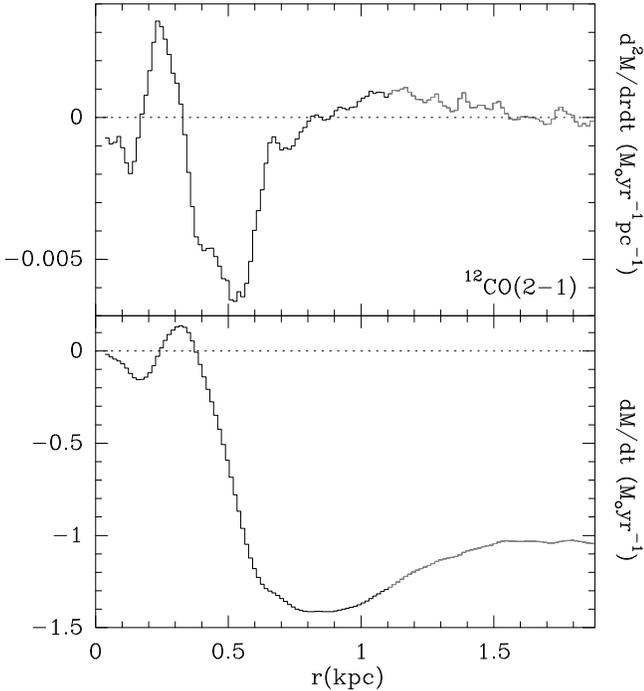}
   \caption{Same as Fig.\ref{fig:mass-radial-HI} but here based on the gas distribution as derived from CO(2--1) data.}
\label{fig:mass-radial-CO21}
 \end{figure}

%%%%%%%%%%%%%%%%%%%%%%%%%%%%%%%%%%%%%%%%%%%%%%%%%%%%%%%%%%%%%%%%%%%%%%%%%%%%%%%%%%%%%%%%%%%%%%%%%%%

\subsubsection{Mass transfer rates in NGC~4579}\label{NGC4579-mass}

In terms of mass inflow/outflow rates, the feeding budget integrated out to a certain radius is clearly negative 
in the outer disk of NGC~4579 ($r>$4~kpc). As shown in Fig.~\ref{fig:mass-radial-HI}, $dM/dt$ goes from virtually zero inflow at $r$\,$\sim$4~kpc to $\sim$--0.3~M$_{\sun}$~yr$^{-1}$ at $r$\,$\sim$10~kpc.
The minimum of $d^2M/dRdt$, indicative of the largest inflow rate per unit length, is reached at $r$\,$\sim$5.5~kpc, a position close to the corotation of the bar. As argued in Sect.~\ref{NGC4579-transfer}, this paradoxical result can be explained if the gas response in the outer disk is decoupled from the bar potential. The inflow rate integrated on these scales ($dM/dt$[4~kpc$<$\,$r$\,$<$10~kpc]\,$\sim$--0.3~M$_{\sun}$~yr$^{-1}$) is a factor $\sim$4 lower than the gas feeding needs implied by the current SFR integrated on this region (SFR$_{outer}$\,$\sim$1.3M$_{\sun}$~yr$^{-1}$, as derived in Sect.~\ref{tracers-SF}). Note, however, that on these scales, the $dM/dt$ estimated from HI is a lower limit to the total inflow rate, as it does not include the contribution from molecular gas at these radii.  

In the inner disk, the integrated mass feeding budget is systematically negative from $r$\,$\sim$2~kpc down to $r$\,$\sim$400~pc. As shown in Figs.~\ref{fig:mass-radial-CO10} and ~\ref{fig:mass-radial-CO21}, $dM/dt$ goes from $\sim$--0.5~M$_{\sun}$~yr$^{-1}$ at the iILR to $\sim$--1.5~M$_{\sun}$~yr$^{-1}$ at the oILR. This is a factor $\sim$5 larger than the corresponding $dM/dt$ inflow rates measured for the outer disk ($r>$4~kpc). The minimum of $d^2M/dRdt$, indicative of the largest inflow rate per unit length, is reached in the vicinity of the iILR.  The inflow rate integrated on these scales ($dM/dt$[400~pc$<$\,$r$\,$<$2~kpc]\,$\sim$--1.5~M$_{\sun}$~yr$^{-1}$) is an order of magnitude higher than the gas feeding rate required to sustain the current SFR in this region (SFR$_{inner}$\,$\sim$0.10~M$_{\sun}$~yr$^{-1}$, as derived in Sect.~\ref{tracers-SF}). The time-scale for gas consumption by star formation is significantly larger than the dynamical time-scale for inflow in the inner disk at present. 

Closer to the AGN ($r$\,$<$200~pc), the integrated mass budget derived from the CO(2--1) map shows negative torques
down to $r$\,$\sim$50~pc. The minimum of $d^2M/dRdt$, indicative of the largest inflow rate per unit length, is reached
inside the UHR of the oval, at $r$\,$\sim$100~pc. The integrated inflow rate measured at $r$\,$\sim$50~pc, the position 
closest to the AGN where $dM/dt$ can be estimated down to the limit imposed by the spatial resolution of our observations, is $dM/dt$\,$\sim$--8$\times$10$^{-3}$~M$_{\sun}$~yr$^{-1}$. This inflow rate can satisfactorily explain 
the AGN activity of NGC~4579: the reported value of $dM/dt$ lies within the range of the mass accretion rates derived from the typical bolometric luminosities in low luminosity AGNs ($\sim$10$^{-2}$ to $\sim$10$^{-5}$~M$_{\sun}$~yr$^{-1}$ from Seyfert to LINERs; see compilation by Jogee~\cite{jog06}).

\section{Summary and conclusions}\label{summary}

In this paper we have studied the efficiency of angular momentum transport in the LINER/Seyfert 1.9 spiral NGC\,4579 using a complete gravity torque map of the disk of the galaxy. Gas flow rates from the outer disk region at  $r$\,$\sim$\,15~kpc down to the inner disk at $\sim$a few 10~pc are derived to search for signatures of secular evolution in the fueling process. The CO maps obtained with the PdBI and the 30m telescope as part of the NUGA survey are used to derive the gravity torque budget in the circumnuclear regions. Gravity torques on the outer disk gas are derived using the HI map obtained with the VLA. The stellar potential is derived from a $K$-band wide field image of the galaxy. We summarize the main results of our study as follows:

\begin{itemize}
\item

By analyzing the distribution and kinematics of the gas and their relation to other tracers of the ISM and the stellar structure we identify signs of dynamical decoupling of several patterns on different spatial scales of the disk. The $K$-band image reveals a stellar bar of diameter D$\sim$12~kpc and moderate strength, as well as a nested weak oval distortion in the inner $r\sim$200~pc of the disk. The bar and the oval are not aligned, a signature of dynamical decoupling. Molecular gas in the inner $r\leq$2~kpc disk is distributed in two spiral arms, an outer ring, and a central lopsided disk. The morphology of the outer disk revealed by HI shows that the neutral gas is currently piling up in a pseudo-ring at $r\geq$4~kpc. The pseudo-ring is formed by two winding spiral arms that are morphologically decoupled from the bar structure; this suggests that the spirals and the bar do not share a common pattern speed.

\item

We have estimated the pattern speeds and principal resonances of the bar and the oval 
in the disk of the galaxy. In the scenario of dynamical decoupling, the oval has a higher pattern speed compared to the primary bar. The gas response in the circumnuclear regions traced by the CO maps follows the expected gas flow pattern induced by the bar potential in the presence of two ILRs at $r\sim$500~pc (iILR) and $r\sim$1.3~kpc (oILR).  We can explain the decoupling of the oval with respect to the large-scale bar, as the corotation of the oval would be close to the oILR of the large-scale bar. The pseudo-ring detected in HI is located inside the bar corotation ($r_{CR}\sim$6~kpc) and close to the predicted position of the UHR of the bar ($r_{UHR}\sim$3.8~kpc). 

\item

The derived gravity torque budget in NGC~4579 shows that the fueling process is at work on different spatial scales in the disk. In the outer disk, the decoupling of the spiral allows the gas to efficiently populate the UHR region inside corotation and thus produce net gas inflow on intermediate scales.  This suggests that the corotation barrier of the bar has been overcome due to secular evolution processes. The gas stops close to the UHR of the bar, where torques become negligible. The gas in the inner disk is efficiently funneled by gravity torques from the oILR down to $r\sim$300~pc.  Closer to the AGN (r$<$200~pc), gas feels negative torques due to the combined action of the large-scale bar and the inner oval. The two $m=2$ modes act in concert to produce net gas inflow down to the spatial resolution of our observations, providing a clear {\it smoking gun} evidence of fueling.

\item 

In the inner disk, the mass feeding budget is systematically negative from $r$\,$\sim$2~kpc down to $r$\,$\sim$400~pc. The inflow rate integrated on these scales ($\sim$--1.5~M$_{\sun}$~yr$^{-1}$) is an order of magnitude higher than the gas feeding rate required to sustain the current SFR in this region.  Closer to the AGN ($r$\,$<$200~pc), the torques are negative down to $r$\,$\sim$50~pc. The integrated inflow rate measured at $r$\,$\sim$50~pc is $\sim$--8$\times$10$^{-3}$~M$_{\sun}$~yr$^{-1}$. This inflow rate can satisfactorily explain the AGN activity of NGC~4579. 

\end{itemize}

NGC~4579 is a representative example of a LLAGN in which a hierarchy of 
mechanisms can act in concert to efficiently transport the gas from the outer 
disk (on kpc scales) down to vicinity of the AGN (on $\sim$a few 10~pc scales) through the 
action of gravity torques. The decoupling of dynamical modes on several 
spatial scales (outer spiral, bar, oval) makes it possible to drain the 
angular momentum of the gas and feed the star formation in the disk, and 
possibly the AGN activity itself. In the last step down to the AGN, viscosity 
can also play a significant role in the fueling process (e.g., see 
discussion in GB05). In particular, we estimate that the efficiency of 
viscosity at a typical distance of $r\sim$50~pc around the central engine of 
NGC~4579\footnote{originally estimated by GB05 at $r\sim$200~pc} may be 
comparable to the efficiency of gravity torques. Both mechanisms can combine 
their efforts and produce sufficient AGN feeding on extremely short dynamical 
time-scales of $\sim$1--2 rotation periods.

\begin{acknowledgements}
	 We acknowledge the IRAM staff from the Plateau de Bure and from 
         Grenoble for carrying out the observations and help provided during the 
	 data reduction. We thank Roberto Neri for fruitful discussions. I.M. acknowledges financial support from the Spanish Ministerio de Ciencia y Tecnolog\'{\i}a (grant AYA2007-62190) and from the Junta de Andaluc\'{\i}a (grant TIC-144). 
      
\end{acknowledgements}


\begin{thebibliography}{}


\bibitem[2004]{and04} Anderson, J.~M., Ulvestad, J.~S., \& Ho, L.~C.\ 2004, \apj, 603, 42 

\bibitem[1992]{ath92} Athanassoula, E.\ 1992, \mnras, 259, 345 

\bibitem[2000]{bak00} Baker, A.~J.\ 2000, 
Ph.D.~Thesis

\bibitem[2003]{bak03} Baker, A.~J., Jogee, S., Sakamoto, K., \& Scoville, N.~Z.\ 2003, Active Galactic Nuclei: From 
Central Engine to Host Galaxy, 290, 479 

\bibitem[1996]{bar96} Barth, A.~J., Reichert, G.~A., Filippenko, A.~V. et al.\ 1996, \aj, 112, 1829 

\bibitem[2001]{bar01} Barth, A.~J., Ho, L.~C., Filippenko, A.~V., Rix, H.-W., \& Sargent, W.~L.~W.\ 2001, \apj, 546, 205 

\bibitem[1978]{boh78} Bohlin, R.~C., Savage, B.~D., \& Drake, J.~F.\ 1978, \apj, 224, 132 

\bibitem[1999]{bok99} B{\"o}ker, T., Calzetti, D., Sparks, W. et al.\ 1999, \apjs, 124, 95 

\bibitem[2007]{boo07} Boone, F., Baker, A.~J., Schinnerer, E. et al.\ 2007, \aap, 471, 113 

\bibitem[2002]{bou02} Bournaud, F., \& Combes, F.\ 2002, \aap, 392, 83 

\bibitem[1992]{bra92} Braine, J., \& Combes, F.\ 1992, \aap, 264, 433 

%\bibitem[1960]{bra60} Brandt, J.~C.\ 1960, \apj, 131, 293 

\bibitem[1996]{but96} Buta, R., \& Combes, F.\ 1996, \fcp, 17, 95 

\bibitem[1993]{can93} Canzian, B.\ 1993, \apj, 414, 487

\bibitem[1989]{car89} Cardelli, J.~A., 
Clayton, G.~C., \& Mathis, J.~S.\ 1989, \apj, 345, 245 

\bibitem[2007]{cas07} Casasola, V., Combes, F., Garc\'{\i}a-Burillo, S. et al.\ 2007, ArXiv e-prints, 712, arXiv:0712.0294 

\bibitem[2006]{che06} Chemin, L., Balkowski, C., Cayatte, V. et al.\ 2006, \mnras, 366, 812 

\bibitem[1988]{com88} Combes, F. 1988, in Galactic and Extragalactic Star Formation, ed. by R.~E. Pudritz \& M.~Fiched, NATO Advanced Science Institutes (ASI) Series C, Volume 232, p.~475
      
\bibitem[2003]{com03} Combes, F. 2003, in Active Galactic Nuclei: from Central Engine to Host Galaxy, ed. by S. Collin, F. Combes \& I. Shlosman. ASP (Astronomical Society of the Pacific), Conference Series, Vol. 290, p. 411.
 
\bibitem[2007]{com07} Combes, F.\ 2007, ArXiv 
e-prints, 709, arXiv:0709.0091 

\bibitem[1981]{com81} Combes, F., \& Sanders, R.~H.\ 1981, \aap, 96, 164
 
\bibitem[2004]{com04} Combes, F., Garc\'{\i}a-Burillo, S., Boone, F. et al.\ 2004, \aap, 414, 857
 
\bibitem[2008]{come08} Comer{\'o}n, S., Knapen, J.~H., Beckman, J.~E., \& Shlosman, I.\ 2008, \aap, 478, 403 


\bibitem[2004]{con04} Contini, M.\ 2004, \mnras, 354, 675

\bibitem[2006]{dai06} Daigle, O., Carignan, C., Amram, P. et al.\ 2006, \mnras, 367, 469 

\bibitem[2004]{dew04} Dewangan, G.~C., Griffiths, R.~E., Di Matteo, T., \& Schurch, N.~J.\ 2004, \apj, 607, 788 

\bibitem[2002]{era02} Eracleous, M., Shields, J.~C., Chartas, G., \& Moran, E.~C.\ 2002, \apj, 565, 108 

\bibitem[2002]{esk02} Eskridge, P.~B., Frogel, J.~A., Pogge, R.~W. et al.\ 2002, \apjs, 143, 73 

\bibitem[2005]{fer05} Ferrarese, L., \& Ford, H.\ 2005, \ssr, 116, 523 

\bibitem[1985]{fil85} Filippenko, A.~V., \& Sargent, W.~L.~W.\ 1985, \apjs, 57, 503 

\bibitem[1993]{gb03} Garc\'{\i}a-Burillo, S., Guelin, M., \& Cernicharo, J.\ 1993, \aap, 274, 123

\bibitem[1998]{gb98} Garc\'{\i}a-Burillo, S., Sempere, M.~J., Combes, F., \& Neri, R.\ 1998, \aap, 333, 864 


% \bibitem[2000]{gb00} Garc\'{\i}a-Burillo, S., Sempere, M.~J., Combes, F. et al. 2000,  A\&A, 363,
% 869
 
\bibitem[2003a]{gb03a} Garc\'{\i}a-Burillo, S., Combes, F., Eckart, A. et al.\ 2003a, in ASP Conf. Ser., Active Galactic Nuclei: from Central Engine to Host Galaxy, ed. by S. Collin, F. Combes,\& I. Shlosman, 423.

\bibitem[2003b]{gb03b} Garc\'{\i}a-Burillo, S., Combes, F., Hunt , L.~K. et al.\ 2003b, \aap, 407, 485 

\bibitem[2005]{gb05} Garc{\'{\i}}a-Burillo, S., Combes, F., Schinnerer, E., Boone, F., \& Hunt, L.~K.\ 2005, \aap, 441, 1011 

\bibitem[2007]{gil07} Gil de Paz, A.; Boissier, S.; Madore, B.~F.; Seibert, M.; Joe, Y.~H. et al.\ 2007, \apjs, 173, 185

\bibitem[2000]{gui00} Guilloteau, S., \& Lucas, R. 2000, in ASP Conf. Ser.: Imaging at Radio through Submillimeter Wavelengths, ed. by J.~G. Mangum \& S. J.~E. Radford, vol. 299

\bibitem[1988]{guh88} Guhathakurta, P., van Gorkom, J.~H., Kotanyi, C.~G., \& Balkowski, C.\ 1988, \aj, 96, 851 

\bibitem[2008a]{haa08a} Haan, S., Schinnerer, E., Mundell, C.~G., Garc{\'{\i}}a-Burillo, S., \& Combes, F.\ 2008a, \aj, 135, 232 

\bibitem[2008b]{haa08b} Haan, S. et al.\ 2008b, \apj, submitted

\bibitem[2001b]{ho01b} Ho, L.~C., \& Ulvestad, J.~S.\ 2001b, \apjs, 133, 77

\bibitem[2001a]{ho01a} Ho, L.~C., Feigelson, E.~D., Townsley, L.~K et al.\ 2001a, \apjl, 549, L51 

\bibitem[1997]{ho97} Ho, L.~C., Filippenko, A.~V., \& Sargent, W.~L.~W.\ 1997, \apjs, 112, 315

 
\bibitem[1995]{hol95} Holtzman, J.~A., Burrows, C.~J., Casertano, S., et al.\ 1995, \pasp, 107, 1065

\bibitem[2006]{hop06} Hopkins, P.~F., \& Hernquist, L.\ 2006, \apjs, 166, 1 

\bibitem[1987]{hum87} Hummel, E., van der Hulst, J.~M., Keel, W.~C., \& Kennicutt, R.~C., Jr.\ 1987, \aaps, 70, 517 

\bibitem[2008]{hun08} Hunt, L.~K., Combes, F., Garc\'{\i}a-Burillo, S. et al.\ 2008, \aap, 482, 133


 %\bibitem[1999]{hun99} Hunt, L.~K., \& Malkan, M.~A. 1999, ApJ, 516, 660
 
% \bibitem[2004]{huntmalkan04} Hunt, L.~K., \& Malkan, M.~A. 2004,
%ApJ, 616, 707

\bibitem[1992]{hut92} Hutchings, J.~B., \& Neff, S.~G.\ 1992, \aj, 104, 1 


\bibitem[2003]{jar03} Jarrett, T.~H., Chester, T., Cutri, R., Schneider, S.~E.,  \&  Huchra, J.~P.\ 2003, \aj, 125, 525
	
%\bibitem[2003]{jar03} Jarrett, T.~H., Chester, T., Cutri, R., Schneider, S.~E., \& Huchra, J.~P., 2003, AJ, 125, 525	

\bibitem[2006]{jog06} Jogee, S.\ 2006, Physics of 
Active Galactic Nuclei at all Scales, 693, 143 


\bibitem[2001]{jog01} Jogee, S., Baker, A.~J., Sakamoto, K. et al. 2001, in ASP Conf. Ser.:The Central Kiloparsec 
of Starbursts and AGN: The La Palma Connection, vol 249, 612
 


% \bibitem[1996]{jun96} Junqueira, S., \& Combes, F. 1996, A\&A, 312, 703 
 
\bibitem[1998]{ken98} Kennicutt, R.~C., Jr.\ 1998, \araa, 36, 189 

\bibitem[1989]{ken89} Kenney, J.~D.~P., \& Young, J.~S.\ 1989, \apj, 344, 171 


\bibitem[2003]{kenn03} Kennicutt, R.~C.,~Jr., Armus, L., Bendo, G. et al. \ 
2003, \pasp, 115, 928


\bibitem[2007]{kin07} King, A.~R., \& Pringle, J.~E.\ 2007, \mnras, 377, L25

\bibitem[2000]{kna00} Knapen, J.~H., Shlosman, I., \& Peletier, R.~F.\ 2000, \apj, 529, 93
 
\bibitem[2003]{kna03} Knapen, J.~H., de Jong, R.~S., Stedman, S., \& Bramich, D.~M.\ 2003, \mnras, 344, 527 

\bibitem[2001]{koh01} Kohno, K., Matsushita, S., Vila-Vilar\'o, B. et al.\ 2001 in ASP Conf. Ser.:The Central 
Kiloparsec of Starbursts and AGN: The La Palma Connection, vol 249, 672
  
\bibitem[2001]{koo01} Koopmann, R.~A., Kenney, J.~D.~P., \& Young, J.\ 2001, \apjs, 135, 125 

\bibitem[2007]{kra07} Krause, M., Fendt, C., \& Neininger, N.\ 2007, \aap, 467, 1037 

\bibitem[2005]{kri05} Krips, M., Eckart, A., Neri, R. et al.\ 2005, \aap, 442, 479 

\bibitem[2007]{kri07} Krips, M., Eckart, A., Krichbaum, T.~P. et al.\ 2007, \aap, 464, 553 

\bibitem[2007]{kun07} Kuno, N., Sato, N., Nakanishi, H. et al.\ 2007, \pasj, 59, 117 

\bibitem[2008]{lin08} Lindt-Krieg, E., Eckart, A., Neri, R. et al.\ 2008, \aap, 479, 377 

%  \bibitem[2004a]{mac04a} Maciejewski, W. 2004a, MNRAS, 354, 883
  
% \bibitem[2004b]{mac04b} Maciejewski, W. 2004b, MNRAS, 354, 892

% \bibitem[2000]{mac00} Maciejewski, W., \& Sparke, L.~S. 2000, MNRAS, 313, 74

% \bibitem[2002]{mac02} Maciejewski, W., Teuben, P.~J., Sparke, L.~S. et al. 2002, MNRAS, 329, 502
 
% \bibitem[1998]{mag98} Magorrian, J., Tremaine, S., Richstone, D. et al. 1998, AJ, 115, 2285
 
%\bibitem[1995]{mao95} Maoz, D., Filippenko, 
%A.~V., Ho, L.~C., Rix, H.-W., Bahcall, J.~N., Schneider, D.~P., \& 
%Macchetto, F.~D.\ 1995, \apj, 440, 91 

\bibitem[1998]{mao98} Maoz, D., Koratkar, A., Shields, J.~C. et al.\ 1998, \aj, 116, 55 

\bibitem[2005]{mao05} Maoz, D., Nagar, N.~M., Falcke, H., Wilson, A.~S.\ 2005, 
\apj, 625, 699

% \bibitem[1993]{mar93} M\'arquez, I., \& Moles, M. 1993, AJ, 105, 2090
 
\bibitem[2000]{mar00} M{\'a}rquez, I., Durret, F., Masegosa, J. et al.\ 2000, \aap, 360, 431 

\bibitem[2004]{mar04} Martini, P. 2004 in Coevolution of Black Holes and Galaxies, ed. by L.~C. Ho, Cambridge University Press, p.~170.
  
% \bibitem[1999]{mar99} Martini, P., \& Pogge, R.~W. 1999, AJ, 118, 2646
 
% \bibitem[2003]{mar03} Martini, P., Regan, M.~W., Mulchaey, J.~S. et al. 2003, ApJ, 589, 774
 
% \bibitem[2004]{mer04} Merloni, A. 2004, MNRAS, 353. 1035	
 
% \bibitem[1995]{mol95} Moles, M., M\'arquez, I., \& P\'erez, E. 1995, ApJ, 438, 604
 
\bibitem[1997]{mul97} Mulchaey, J.~S., \& Regan, M.~W.\ 1997, \apj, 482, L135
 
% \bibitem[2000]{nar00} Narayan, R., Igumenshchev, I.~V., \& Abramowicz, M.~A. 2000, ApJ, 539, 798
 
% \bibitem[1996]{nor96} Norman, C.~A., Sellwood, J.~A., \& Hasan, H. 1996, ApJ, 462 114
 
% \bibitem[2000]{origlia} Origlia, L., \& Leitherer, C. 2000, AJ, 119, 2018


\bibitem[2004]{pah04} Pahre, M.~A., Ashby, 
M.~L.~N., Fazio, G.~G., \& Willner, S.~P.\ 2004, \apjs, 154, 235 

\bibitem[2006]{pee06} Peeples, M.~S., \& Martini, P.\ 2006, \apj, 652, 1097 

\bibitem[1989]{pog89} Pogge, R.~W.\ 1989, \apjs, 71, 433 
\bibitem[2000]{pog00} Pogge, R.~W., Maoz, D., Ho, L.~C., \& Eracleous, M.\ 2000, \apj, 532, 323 


% \bibitem[2000]{per00} P\'erez, E., M\'arquez, I., Marrero, I. et al. 2000, A\&A, 353, 893
  
% \bibitem[1994]{pfe94} Pfenniger, D., \& Combes, F. 1994, A\&A, 285, 94 
 
% \bibitem[2002]{pog02} Pogge, R.~W., \& Martini, P. 2002, ApJ, 569, 624
 
% \bibitem[1981]{pri81} Pringle, J.~E. 1981, ARA\&A, 19, 137

% \bibitem[1994]{quillen} Quillen, A.~C., Frogel, J.~A., \& Gonzalez, R.~A. 1994, ApJ, 437, 162 
  
% \bibitem[1999]{reg99} Regan, M.~W., \& Mulchaey, J.~S. 1999, AJ, 117, 2676
  
\bibitem[2003]{reg03} Regan, M.~W., \& Teuben, P.\ 2003, \apj, 582, 723 

\bibitem[2004]{reg04} Regan, M.~W., \& Teuben, P.~J.\ 2004, \apj, 600, 595 

% \bibitem[2001]{reg01} Regan, M.~W., Thornley, M.~D., Helfer, T.~T. et al. 2001, ApJ, 561, 218
 
\bibitem[1999]{rub99} Rubin, V.~C., Waterman, A.~H., \& Kenney, J.~D.~P.\ 1999, \aj, 118, 236 

  
% \bibitem[2000]{sch00} Schinnerer, E., Eckart, A., Tacconi, L. J. et al. 2000, ApJ, 533, 850

% \bibitem[2001]{sch01} Schmitt, H.~R. 2001, AJ, 122, 2243
 
\bibitem[1988]{sel88} Sellwood, J.~A., \& Sparke, L.~S.\ 1988, \mnras, 231, 25P 

% \bibitem[2004]{sel04} Sellwood, J.~A., \& Shen, J. 2004 in Coevolution of Black Holes and
% Galaxies, from the Carnegie Observatories Centennial Symposia, ed by L.~C. Ho , Cambridge
% University Press, 204.
 
% \bibitem[1989]{shl89} Shlosman, I., Frank, J., Begelman, M.~C. 1989, Nature, 338, 45
 
% \bibitem[1990]{shl90} Shlosman, I., Begelman, M.~C., Frank, J. 1990, Nature, 345, 679
 
% \bibitem[1990]{shu90} Shu, F. H., Tremaine, S., Adams, F. C., \& Ruden, S. P. 1990, ApJ, 358, 49 

\bibitem[2003]{san03} Sanders, D.~B., Mazzarella, J.~M., Kim, D.~-C., Surace, J.~A., \& Soifer, B.~T. \ 2003, \aj, 126, 1607 

\bibitem[1991]{sol91} Solomon, P.~M., \& Barrett, J.~W. 1991 in Dynamics of Galaxies and their Molecular Cloud  Distributions, from the IAU Symp. 146, ed by F.~Combes \& F.~Casoli, Kluwer Academic Publishers, 235.

\bibitem[1982]{sta82} Stauffer, J.~R.\ 1982, \apj, 262, 66 

\bibitem[2000]{ter00} Terashima, Y., Ho, L.~C., Ptak, A.~F.  et al.\ 2000, \apjl, 535, L79

%\bibitem[1988]{tul88} Tully, B. 1988, Nearby Galaxies Catalog (Cambridge University Press,
% Cambridge and New York) 

\bibitem[1988]{tul88} Tully, R.~B., \& Fisher, J.~R.\ 1988, Catalog of Nearby Galaxies, by R.~Brent Tully and 
J.~Richard Fisher, pp.~224.~Cambridge, UK: Cambridge University Press, April 1988


\bibitem[2001]{ulv01} Ulvestad, J.~S., \& Ho, L.~C.\ 2001, \apjl, 562, L133 

\bibitem[2004]{wad04} Wada, K. 2004 in Coevolution of Black Holes and Galaxies, ed. by L.~C. Ho, Cambridge University Press, p.~187

%\bibitem[1998]{wad98} Wada, K., Sakamoto, K., \& Minezaki, T. 1998, ApJ, 494, 236
 	
  
\end{thebibliography}
\end{document}